\documentclass{IEEEtran}
\usepackage{array}
\usepackage{float}
\usepackage{textcomp}
\usepackage{multirow}
\usepackage{amsthm}
\usepackage{amsmath}
\usepackage{amssymb}
\usepackage{graphicx}
\usepackage{units}
\usepackage{amsthm}
\usepackage{amsfonts}
\usepackage{cite}
\usepackage{array}
\usepackage{algorithm}
\usepackage{algorithmic}
\usepackage{subfigure}
\usepackage{stfloats}
\makeatletter

\providecommand{\tabularnewline}{\\}
\floatstyle{ruled}
\newfloat{algorithm}{tbp}{loa}
\providecommand{\algorithmname}{Algorithm}
\floatname{algorithm}{\protect\algorithmname}

\theoremstyle{plain}
\newtheorem{thm}{\protect\theoremname}

\providecommand{\tabularnewline}{\\}
\floatstyle{ruled}
\newfloat{algorithm}{tbp}{loa}
\providecommand{\algorithmname}{Algorithm}
\floatname{algorithm}{\protect\algorithmname}

\DeclareMathOperator{\rank}{rank}
\DeclareMathOperator{\tr}{Tr}

\DeclareMathOperator{\maximize}{maximize}

\DeclareMathOperator{\st}{subject\;to}

\makeatother

\providecommand{\theoremname}{Theorem}

\begin{document}

\title{On the Spectral Efficiency of Full-Duplex Small Cell Wireless Systems }

\author{Dan Nguyen,~\IEEEmembership{Student~Member,~IEEE}, Le-Nam Tran,~\IEEEmembership{Member,~IEEE},
Pekka Pirinen,~\IEEEmembership{Senior Member,~IEEE}, and Matti Latva-aho,~\IEEEmembership{Senior Member,~IEEE}%
\thanks{Copyright (c) 2013 IEEE. Personal use of this material is permitted. However, permission to use this material for any other
purposes must be obtained from the IEEE by sending a request to pubs-permissions@ieee.org.
}%
\thanks{The authors are with the Department of Communications Engineering
and Centre for Wireless Communications, University of Oulu, Finland.
Email: \{vnguyen, ltran, pekka.pirinen, matti.latvaaho\}@ee.oulu.fi. %
}%
\thanks{
The research leading to these results has received funding from the
European Union Seventh Framework Programme (FP7/2007-2013) under grant
agreement n\textdegree{} 316369 \textendash{} project DUPLO and from
the Academy of Finland under grant agreement n\textdegree{} 260755
\textendash{} project Juliet.%
}}

\maketitle
%\vspace{-2cm}
\begin{abstract}
We investigate the spectral efficiency of full-duplex small cell wireless
systems, in which a full-duplex capable base station (BS) is designed
to send/receive data to/from multiple half-duplex users on the same
system resources. The major hurdle for designing such systems is due
to the self-interference at the BS and co-channel interference among
users. Hence, we consider a joint beamformer design to maximize the
spectral efficiency subject to certain power constraints. The design
problem is first formulated as a rank-constrained optimization one,
and the rank relaxation method is then applied. However the relaxed
problem is still nonconvex, and thus optimal solutions are hard to
find. Herein, we propose two provably convergent algorithms to obtain
suboptimal solutions. Based on the concept of the Frank-Wolfe algorithm, we approximate the design problem by a determinant
maximization program in each iteration of the first algorithm. The
second method is built upon the sequential parametric convex approximation
method, which allows us to transform the relaxed problem into a semidefinite
program in each iteration. Extensive numerical experiments under small
cell setups illustrate that the full-duplex system with the proposed
algorithms can achieve a large gain over the half-duplex one. \end{abstract}

\section{Introduction}

The ever growing demand of high data rates and proliferation of a
number of users for wireless services have asked for modern communications
technologies that exploit finite radio resources more efficiently.
Among those, the multiple-input multiple-output (MIMO) communications
technique \cite{Telatar:MIMO:1999} has gradually become a core component
to many wireless communications standards such as LTE \cite{LTE:TS36814}
and WiMAX \cite{802_16_2009}. In the physical layer of wireless communications
networks, MIMO techniques are employed in both downlink and uplink
transmissions. Due to practical limitations on hardware designs, downlink
and uplink channels are currently designed to operate in one dimension
(i.e., either in time or frequency domain). For example, cellular
networks with time division duplex allocate the same frequency band,
but different time slots, to downlink and uplink channels. On the
other hand, cellular networks with frequency division duplex allow
downlink and uplink transmissions to take place in the same time slot,
but over distinct frequencies. Consequently, the radio resources have
not been maximally used in existing wireless communications systems.

Full-duplex transmissions have recently gained significant attention
owing to the potential to further improve or even double the capacity
of conventional half-duplex systems. The benefits of full-duplex systems
are of course brought by allowing the downlink and uplink channels
to function at the same time and frequency \cite{Choi:Fullduplex:achieving:2010,Jain:Fullduplex:Practical:2011,Duarte:Fullduplex:offshelfradios,Duarte:Fullduplex:Experiment,Choi:Fullduplex:beyond,Dan:FD_MIMO:2012,Lee:beamforming:FullDuplexRelay,Day:fullduplexMIMO:AchievableRate,Zheng:Fullduplex:CooperativeCognitive,Aryafar:Fullduplex:MIDU,Dan:Fullduplex:MIMO:2013,Duplo:D1.1,Duplo,Dinesh:FullduplexRadio:Sigcom:2013,Ng:Fullduplex:Relay:MIMO-OFDMA,Day:fullduplexMIMO:Relay:AchievableRate}.
Though the gains of full-duplex systems can be easily foreseen, practical
implementations of such full-duplex systems pose many challenges and
a lot of technical problems still need to be solved before we can
see the first trial deployment on a system level. The crucial barrier
in implementing full-duplex systems resides in the self-interference
(SI) from the transmit antennas to receive antennas at a wireless
transceiver. More explicitly, the radiated power of the downlink channel
interferes with its own desired received signals in the uplink channel.
Clearly, the performance of full-duplex systems depends on the capability
of self-interference cancellation at the transceiver which is limited
in practice. In the past full-duplex transmission was thought infeasible.
This is because the self-interference power, if not efficiently suppressed,
significantly raises the noise floor at receive antennas, exceeding
a limited dynamic range of the analog-to-digital converter (ADC) in
the receiving device \cite{Duarte:Fullduplex:Experiment}.

In recent years, many breakthroughs in hardware design for self-interference
cancellation techniques have been reported, e.g., in \cite{Choi:Fullduplex:achieving:2010,Duarte:Fullduplex:offshelfradios,Jain:Fullduplex:Practical:2011,Dinesh:FullduplexRadio:Sigcom:2013}.
Especially, these studies demonstrate the feasibility of full-duplex
transmission for short to medium range wireless communications. Since
then, several studies focusing on full-duplex communications have
been carried out in a variety of contexts such as point to point MIMO
\cite{Aryafar:Fullduplex:MIDU,Choi:Fullduplex:beyond,Day:fullduplexMIMO:AchievableRate},
MIMO relay \cite{Lee:beamforming:FullDuplexRelay,Ng:Fullduplex:Relay:MIMO-OFDMA,Day:fullduplexMIMO:Relay:AchievableRate},
cognitive radio \cite{Zheng:Fullduplex:CooperativeCognitive}, and
multiuser MIMO systems \cite{Dan:FD_MIMO:2012,Dan:Fullduplex:MIMO:2013}.
With the aim of accelerating full-duplex applications in practical
wireless systems, the full-duplex radios for local access (DUPLO)
project has been funded by the European community's seventh framework
program \cite{Duplo}. As a first step, the first deliverable of the
DUPLO project has identified several potential deployment scenarios
that may benefit from full-duplex communications \cite{Duplo:D1.1}.
Among others, small cell wireless communications systems are selected
as one of the important research frameworks. In fact, small cell systems
are considered to be especially suitable for deployment of full-duplex
technology due to low transmit powers, short transmission distances
and low mobility.

What is missing in \cite{Duplo:D1.1} is further studies that evaluate
the actual gain of the full-duplex transmission for some reference
systems. The goal of this paper is to fill this gap. Particularly,
we consider a scenario where a full-duplex capable base station (BS)
communicates with half-duplex users in both directions at the same
time slot over the same frequency band. It is now well known that
the optimal transmit strategy for downlink channels is achieved by
dirty paper coding \cite{Weingarten:CapacityRegion:MU_MIMO:2006},
but it requires high complexity to implement. Thus, we adopt a linear beamforming technique for the downlink transmission in this
paper, which has been widely used in the literature, e.g., in \cite{Bengtsson:2589,mats:convexbasedBeamforming,Nam:WSRM:SOCP:2012}.
For uplink channels, the optimal nonlinear multiuser detection scheme
based on minimum mean square error and successive interference cancellation
(MMSE-SIC) \cite{Tse:WCOM:2005} is chosen in this paper. For the
considered full-duplex system, the problem of beamformer design becomes
more challenging since there still exists a small, but not negligible,
amount of the self-interference between the transmit and receive antennas
at the BS even after an advanced SI cancellation technique is applied.
We note also that the SI level increases with the transmit power for
any SI cancellation technique. Moreover, the difficulty of the design
problem is increased further by the co-channel interference (CCI)
caused by the users in the uplink channel to those in the downlink
channel.%
\footnote{Through out the paper, the co-channel interference refers to the interference
that users in the uplink cause for those in the downlink channel,
not the mutual interference among users in the downlink channel. %
} By this very nature, a joint design of the downlink and uplink transmissions
would offer the best solution. One of the first attempts to investigate
the potential gain of full-duplex systems has been made in our earlier
work of \cite{Dan:FD_MIMO:2012,Dan:Fullduplex:MIMO:2013}. However
the CCI is not taken into account and several system parameters were
ideally assumed therein. These practical considerations are carefully
examined in this paper.

We are concerned with the problem of joint beamformer design to maximize
the spectral efficiency (SE) under some power constraints. To this
end, the total SE maximization (SEMax) problem is first formulated
as a rank-constrained optimization problem for which it is difficult
to find globally optimal solutions in general. The standard method
of rank relaxation is then applied to arrive at a relaxed problem,
which is still nonconvex. After solving the relaxed problem, the randomization
technique presented in \cite{Sidiropoulos:MulticastBeamforming:2006}
is employed to find the beamformers for the original design problem.
We note that the rank relaxation technique, commonly known as semidefinite
relaxation (SDR) method under various contexts, is widely used to
solve the problem of linear precoder design in MIMO downlink channels,
e.g., in \cite{Sidiropoulos:MulticastBeamforming:2006,Luo:SDR:2010,Wiesel:ZeroForcing:GeneralInverse:2008,Nameamdesign:ZFDPC:2012,Nam:SZFDPC:2012,mats:convexbasedBeamforming}.
Very often, the relaxed problems in those cases are convex and general
convex program solvers can be called upon to find the solutions. Moreover,
in some special cases, the rank relaxation is proved to be tight \cite{Bengtsson:2589,Wiesel:ZeroForcing:GeneralInverse:2008,Nam:SZFDPC:2012}.
The same property unfortunately does not carry over into our case.

To tackle the nonconvexity of the relaxed problem, we propose two
iterative local optimization algorithms. The first proposed algorithm
is a direct result of exploiting the `difference of convex' (D.C)
structure of the relaxed problem. To be specific, based on the idea
of the Frank-Wolfe (FW) algorithm \cite{FrankWolfe:1956}, we arrive
at a determinant maximization (MAXDET) program at each iteration.
The second design approach involves some transformations before invoking
the framework of sequential parametric convex approximation (SPCA)
method \cite{Beck:SCA:2010}, which has proven to be an effective
tool for numerical solutions of nonconvex optimization problems \cite{Nam:WSRM:SOCP:2012,Nam:SZF:2012,Beck:SCA:2010}.
In particular, we are able to approximate the relaxed problem as a
semidefinite program (SDP) at each iteration in the second iterative
algorithm. While the first design algorithm sticks to MAXDET problem
solvers, the second one offers more flexibility in choosing optimization
software and can take advantage of many state-of-the-art SDP solvers.
Additionally, since there is no (even rough) way to estimate beforehand
which algorithm is better than the other for a given set of channel
realizations, the two iterative algorithms can be implemented in a
concurrent manner and a solution is obtained when one of them terminates.
Alternatively, we run the two algorithms in parallel until they converge,
and then select the better solution. The numerical results on the
SE and computational complexity of the two methods are given in Section
\ref{sec:numerical results}.

Full-duplex transmission, if successfully implemented, is clearly
expected to improve the spectral efficiency of wireless communications
systems. However, a quantitative answer on the potential
gains for some particular scenarios is still missing. For this purpose,
the proposed algorithms are used to evaluate the performance of the
full-duplex system of consideration under the 3GPP LTE specifications
for a small cell system. The numerical experiments demonstrate that
small cell full-duplex transmissions are superior to the conventional
half-duplex one as long as the self-interference power is efficiently
canceled.

The rest of the paper is organized as follows. The full-duplex system
model and problem formulation are presented in Section \ref{sec:SM_PF}.
In Section \ref{sec:designs}, we describe the proposed iterative
beamformer designs. The SE performance of the considered full-duplex
transmission is numerically compared to the conventional half-duplex
one in Section IV. Finally, the paper concludes with
future work in Section \ref{sec:Conclusions}.

\textit{Notation}: We use standard notations in this paper. Namely,
bold lower and upper case letters represent vectors and matrices,
respectively; $\mathbf{H}^{H}$ and $\mathbf{H}^{T}$ are Hermitian
and standard transpose of $\mathbf{H}$, respectively; $\tr(\mathbf{H})$
and $|\mathbf{H}|$ are the trace and determinant of $\mathbf{H}$,
respectively; $\mathbf{H}\succeq\mathbf{0}$ means that $\mathbf{H}$
is a positive semidefinite matrix; $\rank(\mathbf{H})$ is rank of
$\mathbf{H}$; $\nabla_{\mathbf{X}}\, f(\mathbf{X})$ is the gradient
of $f(\mathbf{X})$; $E(\cdot)$ denotes the expectation operator.

\section{\label{sec:SM_PF}System Model and Problem Formulation}
\begin{figure}
\centering\includegraphics[width=0.95\columnwidth]{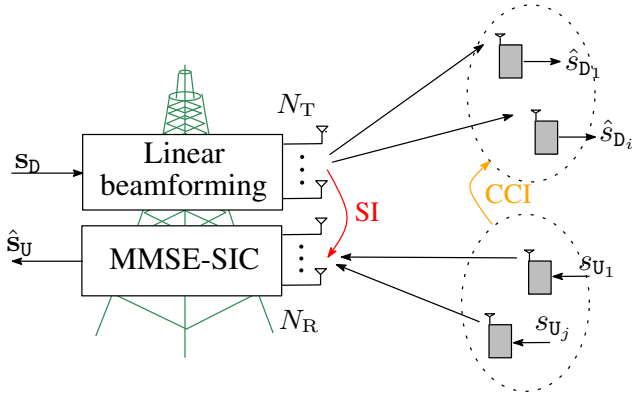}
\caption{A small cell full-duplex wireless communications system. The number
of transmit and receive antennas at the BS is $N_{\mathtt{T}}$ and
$N_{\mathtt{R}}$, respectively. Linear beamforming is adopted for
the downlink channel, while MMSE-SIC for the uplink channel. In the
figure, SI and CCI mean self-interference and co-channel interference,
respectively.}
\label{fig:FD:SystemModel}
\end{figure}

We consider a small cell full-duplex wireless communications system
in which a full-duplex capable BS is designed to communicate with
$K_{\mathtt{D}}$ single-antenna users in the downlink channel and
$K_{\mathtt{U}}$ single-antenna users in the uplink channel at the
same time over the same frequency band, as depicted in Fig. \ref{fig:FD:SystemModel}.
Throughout the paper, the notations $\mathtt{D}_{i}$ and $\mathtt{U}_{j}$
refer to the $i$th and $j$th user in the downlink and uplink channels,
respectively. The total number of antennas at the full-duplex BS is
$N=N_{\mathtt{T}}+N_{\mathtt{R}}$, of which $N_{\mathtt{T}}$ transmit
antennas are used for data transmissions in the downlink channel and
$N_{\mathtt{R}}$ receive antennas are dedicated to receiving data
in the uplink channel. We further assume that the channels are flat
fading and channel state information (CSI) is perfectly known at both
the BS and users.

First, in the downlink channel, let $s_{\mathtt{D}_{i}}$ be the transmitted
data symbol for $\mathtt{D}_{i}$, which is normalized to $E(\bigl|s_{\mathtt{D}_{i}}\bigr|^{2})=1$.
For linear beamforming, the data symbol $s_{\mathtt{D}_{i}}$ is multiplied
by the beamforming vector $\mathbf{w}_{\mathtt{D}_{i}}\in\mathbb{C}^{N_{\mathtt{T}}\times1}$
before transmission, and the received signal of user $\mathtt{D}_{i}$
is given by
\begin{equation}
y_{\mathtt{D}_{i}}=\mathbf{h}_{\mathtt{D}_{i}}^{H}\mathbf{w}_{\mathtt{D}_{i}}s_{\mathtt{D}_{i}}+\underbrace{\sum_{k\neq i}^{K_{\mathtt{D}}}\mathbf{h}_{\mathtt{D}_{i}}^{H}\mathbf{w}_{\mathtt{D}_{k}}s_{\mathtt{D}_{k}}}_{\textrm{MUI}}+\underbrace{\sum_{j=1}^{K_{\mathtt{U}}}g_{ji}s_{\mathtt{U}_{j}}}_{\textrm{CCI}}+n_{\mathtt{D}_{i}}\label{eq:DL:SignalModel}
\end{equation}
where $\mathbf{h}_{\mathtt{D}_{i}}$ is the $N_{\mathtt{T}}\times1$
complex channel vector from the BS to user $\mathtt{D}_{i}$, $g_{ji}$
is the complex channel coefficient from $\mathtt{U}_{j}$ to $\mathtt{D}_{i}$,
$s_{\mathtt{U}_{j}}$ is the data symbol transmitted by $\mathtt{U}_{j}$
in the uplink direction, and $n_{\mathtt{D}_{i}}\sim\mathcal{CN}(0,\sigma_{n}^{2})$
is background noise assumed to be additive white Gaussian (AWGN).
In \eqref{eq:DL:SignalModel}, the first and second summations represent
multiuser interference (MUI) in the downlink channel and co-channel
interference (CCI) from the uplink to the downlink channels, respectively.
The received signal to interference plus noise ratio (SINR) of user
$\mathtt{D}_{i}$ can be written as
\begin{equation}
\begin{array}{rl}
\gamma_{\mathtt{D}_{i}} & =\dfrac{\bigl|\mathbf{h}_{\mathtt{D}_{i}}^{H}\mathbf{w}_{\mathtt{D}_{i}}\bigr|^{2}}{\sigma_{n}^{2}+\sum_{k\neq i}^{K_{\mathtt{D}}}\bigl|\mathbf{h}_{\mathtt{D}_{i}}^{H}\mathbf{w}_{\mathtt{D}_{k}}\bigr|^{2}+\sum_{j=1}^{K_{\mathtt{U}}}q_{\mathtt{U}_{j}}\bigl|g_{ji}\bigr|^{2}}\\
 & =\dfrac{\mathbf{h}_{\mathtt{D}_{i}}^{H}\mathbf{Q}_{\mathtt{D}_{i}}\mathbf{h}_{\mathtt{D}_{i}}}{\sigma_{n}^{2}+\sum_{k\neq i}^{K_{\mathtt{D}}}\mathbf{h}_{\mathtt{D}_{i}}^{H}\mathbf{Q}_{\mathtt{D}_{k}}\mathbf{h}_{\mathtt{D}_{i}}+\sum_{j=1}^{K_{\mathtt{U}}}q_{\mathtt{U}_{j}}\bigl|g_{ji}\bigr|^{2}}
\end{array}\label{eq:DL:SINR}
\end{equation}
where $E(|s_{\mathtt{U}_{j}}|^{2})=q_{\mathtt{U}_{j}}$, $j=1,...,K_{\mathtt{U}}$,
is power loading for user $\mathtt{U}_{j}$ in the uplink direction;
$\mathbf{Q}_{\mathtt{D}_{i}}=\mathbf{w}_{\mathtt{D}_{i}}\mathbf{w}_{\mathtt{D}_{i}}^{H}$,
and $\rank(\mathbf{Q}_{\mathtt{D}_{i}})=1$. Then, spectral efficiency
in the downlink direction is given by %
\footnote{We use natural logarithm for the sake of mathematical convenience.
However, the SE is calculated with logarithm to base $2$ in the numerical
result section.%
} \begin{IEEEeqnarray}{rCl}\IEEEyesnumber\label{eq:DL:SumRate:general} R_{\mathtt{D}} & = & \sum_{i=1}^{K_{\mathtt{D}}}\log(1+\gamma_{\mathtt{D}_{i}})\label{eq:DL:SumRateforSDP}\IEEEyessubnumber\\  & = & \sum_{i=1}^{K_{\mathtt{D}}}\log\left(\frac{\sigma_{n}^{2}+\displaystyle \sum_{k=1}^{K_{\mathtt{D}}}\mathbf{h}_{\mathtt{D}_{i}}^{H}\mathbf{Q}_{\mathtt{D}_{k}}\mathbf{h}_{\mathtt{D}_{i}}+\sum_{j=1}^{K_{\mathtt{U}}}q_{\mathtt{U}_{j}}\bigl|g_{ji}\bigr|^{2}}{\sigma_{n}^{2}+\displaystyle \sum_{k\neq i}^{K_{\mathtt{D}}}\mathbf{h}_{\mathtt{D}_{i}}^{H}\mathbf{Q}_{\mathtt{D}_{k}}\mathbf{h}_{\mathtt{D}_{i}}+\sum_{j=1}^{K_{\mathtt{U}}}q_{\mathtt{U}_{j}}\bigl|g_{ji}\bigr|^{2}}\right).\IEEEeqnarraynumspace\IEEEyessubnumber\label{eq:DL:SumRate} \end{IEEEeqnarray}

Next, for the uplink transmission, we can express the received signal
vector at the full-duplex BS as
\begin{equation}
\mathbf{y}_{\mathtt{U}}=\sum_{j=1}^{K_{\mathtt{U}}}\mathbf{h}_{\mathtt{U}_{j}}s_{\mathtt{U}_{j}}+\underbrace{\sum_{i=1}^{K_{\mathtt{D}}}\mathbf{H}_{\mathtt{SI}}\mathbf{w}_{\mathtt{D}_{i}}s_{\mathtt{D}_{i}}}_{\textrm{self-interference}}+\mathbf{n}_{\mathtt{U}}\label{eq:UL:SignalModel}
\end{equation}
where $\mathbf{h}_{\mathtt{U}_{j}}\in\mathbb{C}^{N_{\mathtt{R}}\times1}$
is the complex channel vector from the BS to $\mathtt{U}_{j}$ and
$\mathbf{n}_{\mathtt{U}}\sim\mathcal{CN}(0,\sigma_{n}^{2}\mathbf{I}_{N_{\mathtt{R}}})$.
The matrix $\mathbf{H}_{\mathtt{SI}}$ is called the self-interference
channel from the transmit antennas to the receive antennas at the
full-duplex BS, in which the values of its entries are determined
by the capability of the advanced SI cancellation techniques. In this
case, by treating the self-interference as background noise and applying
the MMSE-SIC decoder, we can write the received SINR of  $\mathtt{U}_{j}$
as \cite{Tse:WCOM:2005}
\begin{equation}
\gamma_{\mathtt{U}_{j}}=q_{\mathtt{U}_{j}}\mathbf{h}_{\mathtt{U}_{j}}^{H}\Bigl(\sigma_{n}^{2}\mathbf{I}+\sum_{m>j}^{K_{\mathtt{U}}}q_{\mathtt{U}_{m}}\mathbf{h}_{\mathtt{U}_{m}}\mathbf{h}_{\mathtt{U}_{m}}^{H}+\sum_{i=1}^{K_{\mathtt{D}}}\mathbf{H}_{\mathtt{SI}}\mathbf{Q}_{\mathtt{D}_{i}}\mathbf{H}_{\mathtt{SI}}^{H}\Bigr)^{-1}\mathbf{h}_{\mathtt{U}_{j}}\label{eq:UL:SINR}
\end{equation}
where we have assumed a decoding order from $1$ to $K_{\mathtt{U}}$.
Consequently, the achievable SE of the uplink channel is given by
\cite{Tse:WCOM:2005}  \begin{IEEEeqnarray}{rCl}\IEEEyesnumber\label{eq:UL:SumRate} R_{\mathtt{U}} & = & \sum_{j=1}^{K_{\mathtt{U}}}\log(1+\gamma_{\mathtt{U}_{j}})\IEEEyessubnumber\label{eq:UL:SumRate:a}\\  & = & \sum_{j=1}^{K_{\mathtt{U}}}\log\Bigl(1+q_{\mathtt{U}_{j}}\mathbf{h}_{\mathtt{U}_{j}}^{H}\Bigl(\sigma_{n}^{2}\mathbf{I}+\sum_{m>j}^{K_{\mathtt{U}}}q_{\mathtt{U}_{m}}\mathbf{h}_{\mathtt{U}_{m}}\mathbf{h}_{\mathtt{U}_{m}}^{H}\nonumber \\  &  & +\sum_{i=1}^{K_{\mathtt{D}}}\mathbf{H}_{\mathtt{SI}}\mathbf{Q}_{\mathtt{D}_{i}}\mathbf{H}_{\mathtt{SI}}^{H}\Bigr)^{-1}\mathbf{h}_{\mathtt{U}_{j}}\Bigr)\IEEEyessubnumber\label{eq:UL:SumRate:a1}\\  & = & \log\dfrac{\Bigl|\sigma_{n}^{2}\mathbf{I}+{\displaystyle \sum_{i=1}^{K_{\mathtt{D}}}}\mathbf{H}_{\mathtt{SI}}\mathbf{Q}_{\mathtt{D}_{i}}\mathbf{H}_{\mathtt{SI}}^{H}+{\displaystyle \sum_{j=1}^{K_{\mathtt{U}}}}q_{\mathtt{U}_{j}}\mathbf{h}_{\mathtt{U}_{j}}\mathbf{h}_{\mathtt{U}_{j}}^{H}\Bigr|}{\Bigl|\sigma_{n}^{2}\mathbf{I}+{\displaystyle \sum_{i=1}^{K_{\mathtt{D}}}}\mathbf{H}_{\mathtt{SI}}\mathbf{Q}_{\mathtt{D}_{i}}\mathbf{H}_{\mathtt{SI}}^{H}\Bigr|}.\IEEEyessubnumber\label{eq:UL:SumRate:b} \end{IEEEeqnarray}

From \eqref{eq:DL:SignalModel} and \eqref{eq:UL:SignalModel}, we
observe that the downlink and uplink transmissions are coupled by
the CCI and self-interference. This problem greatly impacts the performance
of the system of interest. Herein, our main purpose is to jointly
design beamformers so that the total system spectral efficiency is
maximized under the sum transmit power constraint in the downlink
channel and per-user power constraints in the uplink one. Specifically,
the total SEMax problem is formulated
as a rank-constrained optimization one as \begin{subequations}\label{eq:Original:SE}
\begin{eqnarray}
\underset{\{\mathbf{Q}_{\mathtt{D}_{i}}\},\{q_{\mathtt{U}_{j}}\}}{\maximize} &  & R_{\mathtt{D}}+R_{\mathtt{U}}\label{eq:SE:obj}\\
\st &  & 0\leq q_{\mathtt{U}_{j}}\leq\overline{q}_{\mathtt{U}_{j}},\forall j=1,\ldots,K_{\mathtt{U}},\label{eq:SE:c1}\\
 &  & \sum_{i=1}^{K_{\mathtt{D}}}\tr(\mathbf{Q}_{\mathtt{D}_{i}})\leq P_{\mathtt{BS}},\label{eq:SE:c2}\\
 &  & \mathbf{Q}_{\mathtt{D}_{i}}\succeq0,\forall i=1,\ldots,K_{\mathtt{D}},\label{eq:SE:c3}\\
 &  & \rank(\mathbf{Q}_{\mathtt{D}_{i}})=1,\forall i=1,\ldots,K_{\mathtt{D}}\label{eq:SE:c4}
\end{eqnarray}
\end{subequations} where $P_{\mathtt{BS}}$ is the maximum power
at the BS and $\overline{q}_{\mathtt{U}_{j}}$ is the power constraint
at each user in the uplink channel. Clearly, problem \eqref{eq:Original:SE}
is a nonconvex program, which is difficult to solve optimally in general.
We also note that a simplified problem of \eqref{eq:Original:SE},
in which $q_{\mathtt{U}_{j}}$ and $R_{\mathtt{U}}$ are omitted (i.e.,
the SEMax problem for the downlink channel itself), was proved to
be NP-hard \cite{Luo:SpectrumManagement:2008}. Thus, we conjecture
that the NP-hardness is carried over into our problem. Towards a tractable
solution, we first apply the relaxation method to obtain a relaxed
problem of \eqref{eq:Original:SE} by dropping the rank-1 constraints
\eqref{eq:SE:c4}. Then, two efficient iterative algorithms proposed
to solve the resulting problem are presented in the next section.

\section{\label{sec:designs}Proposed Beamformer Designs }

Note that the relaxed problem of \eqref{eq:Original:SE} after dropping
the rank constraints is still nonconvex. Thus, computing its globally
optimal solution is difficult and very computationally expensive in
general. To the best of our knowledge, finding an optimal solution
to the nonconvex problems similar to \eqref{eq:Original:SE} is still
an open problem. In this section, we present two reformulations of
the relaxed problem, based on which two iterative algorithms of different
level of complexity are developed.

\subsection{\label{sub:Log-Approximation}Iterative MAXDET-based Algorithm }

The first beamforming algorithm is built upon an observation that
the SE of the system at hand is a difference of two concave functions.
Indeed, from \eqref{eq:DL:SumRate} and \eqref{eq:UL:SumRate:b},
we can write $R_{\mathtt{D}}+R_{\mathtt{U}}=h(\mathbf{Q},\mathbf{q})-g(\mathbf{Q},\mathbf{q})$,
where
\begin{multline}
h(\mathbf{Q},\mathbf{q})\triangleq\log\Bigl|\sigma_{n}^{2}\mathbf{I}+\sum_{i=1}^{K_{\mathtt{D}}}\mathbf{H}_{\mathtt{SI}}\mathbf{Q}_{\mathtt{D}_{i}}\mathbf{H}_{\mathtt{SI}}^{H}+\sum_{j=1}^{K_{\mathtt{U}}}q_{\mathtt{U}_{j}}\mathbf{h}_{\mathtt{U}_{j}}\mathbf{h}_{\mathtt{U}_{j}}^{H}\Bigr|\\
+\sum_{i=1}^{K_{\mathtt{D}}}\log\Bigl(\sigma_{n}^{2}+\sum_{k=1}^{K_{\mathtt{D}}}\mathbf{h}_{\mathtt{D}_{i}}^{H}\mathbf{Q}_{\mathtt{D}_{k}}\mathbf{h}_{\mathtt{D}_{i}}+\sum_{j=1}^{K_{\mathtt{U}}}q_{\mathtt{U}_{j}}|g_{ji}|^{2}\Bigr),
\end{multline}
\begin{eqnarray}
g(\mathbf{Q},\mathbf{q}) & \triangleq & \sum_{i=1}^{K_{\mathtt{D}}}\log\Bigl(\sigma_{n}^{2}+\sum_{k\neq i}^{K_{\mathtt{D}}}\mathbf{h}_{\mathtt{D}_{i}}^{H}\mathbf{Q}_{\mathtt{D}_{k}}\mathbf{h}_{\mathtt{D}_{i}}+\sum_{j=1}^{K_{\mathtt{U}}}q_{\mathtt{U}_{j}}|g_{ji}|^{2}\Bigr)\nonumber \\
 &  & +\log\Bigl|\sigma_{n}^{2}\mathbf{I}+\sum_{i=1}^{K_{\mathtt{D}}}\mathbf{H}_{\mathtt{SI}}\mathbf{Q}_{\mathtt{D}_{i}}\mathbf{H}_{\mathtt{SI}}^{H}\Bigr|
\end{eqnarray}
and $\mathbf{Q}$ and $\mathbf{q}$ are the symbolic notations that
denote the set of design variables $\{\mathbf{Q}_{\mathtt{D}_{i}}\}$
and $\{q_{\mathtt{U}_{j}}\}$, respectively. It should be noted that
the functions $h(\mathbf{Q},\mathbf{q})$ and $g(\mathbf{Q},\mathbf{q})$
are jointly concave with respect to $\mathbf{Q}$ and $\mathbf{q}$
\cite{Boyd:ConvexOpt:2004}. Borrowing the concept of the FW method
which considers a linear approximation of the objective function and
searches for a direction that improves the objective, we now present
the first joint design algorithm to find $\mathbf{Q}$ and $\mathbf{q}$.
First, the relaxed problem is reformulated as
\begin{equation}
\begin{array}{rl}
\underset{{\scriptstyle \mathbf{Q},\mathbf{q}}}{\maximize} & {\displaystyle h(\mathbf{Q},\mathbf{q})-g(\mathbf{Q},\mathbf{q})}\\
\st & \eqref{eq:SE:c1},\;\eqref{eq:SE:c2},\;\eqref{eq:SE:c3}.
\end{array}\label{eq:DCform}
\end{equation}
Since the constraints \eqref{eq:SE:c1}-\eqref{eq:SE:c3} are convex,
the difficulty in solving \eqref{eq:DCform} lies in the  component
$-g(\mathbf{Q},\mathbf{q})$. Suppose the value of $(\mathbf{Q},\mathbf{q})$
at iteration $n$ is denoted by $(\mathbf{Q}^{(n)},\mathbf{q}^{(n)})$.
To increase the objective in the next iteration we replace $g(\mathbf{Q},\mathbf{q})$
by its affine majorization at a neighborhood of $(\mathbf{Q}^{(n)},\mathbf{q}^{(n)})$.
Since $g(\mathbf{Q},\mathbf{q})$ is concave and differentiable on
the considered domain $\{\mathbf{Q}_{\mathtt{D}_{i}},q_{\mathtt{U}_{j}}:\mathbf{Q}_{\mathtt{D}_{i}}\succeq0,q_{\mathtt{U}_{j}}\geq0\}$,
one can easily find an affine majorization as a first order approximation
as \cite{Boyd:ConvexOpt:2004}
\begin{multline}
g^{(n)}(\mathbf{Q},\mathbf{q})=g(\mathbf{Q}^{(n)},\mathbf{q}^{(n)})+\sum_{i=1}^{K_{\mathtt{D}}}\sum_{k=1,k\neq i}^{K_{\mathtt{D}}}\Bigl[\bigl(\vartheta_{\mathtt{D}_{i}}^{(n)}\bigr)^{-1}\mathbf{h}_{\mathtt{D}_{i}}^{H}\\
\bigl(\mathbf{Q}_{\mathtt{D}_{k}}-\mathbf{Q}_{\mathtt{D}_{k}}^{(n)}\bigr)\mathbf{h}_{\mathtt{D}_{i}}\Bigr]+\sum_{i=1}^{K_{\mathtt{D}}}\sum_{j=1}^{K_{\mathtt{U}}}\bigl(\vartheta_{\mathtt{D}_{i}}^{(n)}\bigr)^{-1}|g_{ji}|^{2}\bigl(q_{\mathtt{U}_{j}}-q_{\mathtt{U}_{j}}^{(n)}\bigr)\\
+\sum_{i=1}^{K_{\mathtt{D}}}\tr\Bigl[\mathbf{H}_{\mathtt{SI}}^{H}\bigl(\boldsymbol{\Theta}^{(n)}\bigr)^{-1}\mathbf{H}_{\mathtt{SI}}\bigl(\mathbf{Q}_{\mathtt{D}_{i}}-\mathbf{Q}_{\mathtt{D}_{i}}^{(n)}\bigr)\Bigr]\label{eq:affine}
\end{multline}
 where $\vartheta_{\mathtt{D}_{i}}^{(n)}$ and $\boldsymbol{\Theta}^{(n)}$
are defined as
\begin{eqnarray}
\vartheta_{\mathtt{D}_{i}}^{(n)} & = & \sigma_{n}^{2}+\sum_{m\neq i}^{K_{\mathtt{D}}}\mathbf{h}_{\mathtt{D}_{i}}^{H}\mathbf{Q}_{\mathtt{D}_{m}}^{(n)}\mathbf{h}_{\mathtt{D}_{i}}+\sum_{l=1}^{K_{\mathtt{U}}}q_{\mathtt{U}_{l}}^{(n)}|g_{li}|^{2},\\
\boldsymbol{\Theta}^{(n)} & = & \sigma_{n}^{2}\mathbf{I}+\sum_{j=1}^{K_{\mathtt{D}}}\mathbf{H}_{\mathtt{SI}}\mathbf{Q}_{\mathtt{D}_{j}}^{(n)}\mathbf{H}_{\mathtt{SI}}^{H}.
\end{eqnarray}
To derive \eqref{eq:affine}, we have used the fact $\nabla_{\mathbf{X}}\,\log|\mathbf{I}+\mathbf{A}\mathbf{X}\mathbf{A}^{H}|=\mathbf{A}^{H}(\mathbf{I}+\mathbf{A}\mathbf{X}\mathbf{A}^{H})^{-1}\mathbf{A}$,
$\nabla_{x}\,\log(1+ax)=a(1+ax)^{-1}$, the inner product of two semidefinite
matrices $\mathbf{X}\succeq\mathbf{0}$ and $\mathbf{Y}\succeq\mathbf{0}$
is $\tr(\mathbf{XY})$, and the inner product of two vector is $\mathbf{x}^{H}\mathbf{y}$
\cite{Dattorro:ConvexOpt:2011}. Now, we approximate problem \eqref{eq:DCform}
at iteration $n+1$ by a convex program given by
\begin{equation}
\begin{array}{rl}
\underset{{\scriptstyle \mathbf{Q},\mathbf{q}}}{\maximize} & h(\mathbf{Q},\mathbf{q})-g^{(n)}(\mathbf{Q},\mathbf{q})\\
\st & \eqref{eq:SE:c1},\;\eqref{eq:SE:c2},\;\eqref{eq:SE:c3}.
\end{array}\label{eq:MAXDET}
\end{equation}
The objective in \eqref{eq:MAXDET} is in fact a lower bound of the
SE of the full-duplex system. We note that problem \eqref{eq:MAXDET}
is a MAXDET program, and hence the name of the first algorithm. Let
$(\mathbf{Q}^{\star},\mathbf{q}^{\star})$ be the optimal value of
$(\mathbf{Q},\mathbf{q})$ in \eqref{eq:MAXDET}. Then we update $(\mathbf{Q}^{(n+1)},\mathbf{q}^{(n+1)}):=(\mathbf{Q}^{\star},\mathbf{q}^{\star})$.
In this way, the design variables are iteratively updated and the
lower bound of the SE increases after every iteration. Since the SE
is bounded above due to the power constraints \eqref{eq:SE:c1} and
\eqref{eq:SE:c2}, the iterative procedure is guaranteed to converge.
The iterative MAXDET-based algorithm is outlined in Algorithm \ref{algo:iterativeMAXDET}.
The convergence properties of Algorithm \ref{algo:iterativeMAXDET}
are stated in Theorem \ref{thm:converge}. %which is shown after Algorithm
%\ref{algo:IterativeSDP}.

\begin{algorithm}
\caption{Iterative MAXDET-based algorithm.}
\label{algo:iterativeMAXDET}\begin{algorithmic}[1]
\renewcommand{\algorithmicrequire}{\textbf{Initialization:}} \REQUIRE
\STATE Generate initial values for $\mathbf{Q}_{\mathtt{D}_{i}}^{(0)}$
for $i=1,2,\ldots,K_{\mathtt{D}}$ and $q_{\mathtt{U}_{j}}^{(0)}$
for $i=1,2,\ldots,K_{\mathtt{U}}$.
\STATE Set $n:=0$.
\renewcommand{\algorithmicrequire}{\textbf{Iterative procedure:}}
\REQUIRE
\REPEAT
\STATE Solve \eqref{eq:MAXDET} and denote the optimal solutions
as $(\mathbf{Q}^{\star},\mathbf{q}^{\star})$.
\STATE Update: $\mathbf{Q}_{\mathtt{D}_{i}}^{(n+1)}:=\mathbf{Q}_{\mathtt{D}_{i}}^{\star}$;
and $q_{\mathtt{U}_{j}}^{(n+1)}:=q_{\mathtt{U}_{j}}^{\star}$ .
\STATE Set $n:=n+1$.
\UNTIL Convergence.
\renewcommand{\algorithmicrequire}{\textbf{Finalization:}}
\REQUIRE
\STATE Perform randomization to extract a rank-1 solution if required.
\label{randomization}
\end{algorithmic}
\end{algorithm}
An important point to note here is that the iterative procedure in
Algorithm \ref{algo:iterativeMAXDET} possibly returns a locally optimal
solution to a relaxed problem of \eqref{eq:Original:SE} at convergence.
Obviously, if $\rank(\mathbf{Q}_{\mathtt{D}_{i}}^{\star})=1$, then
$\mathbf{Q}_{\mathtt{D}_{i}}^{\star}$ is also feasible to \eqref{eq:Original:SE}
and the beamformer for $\mathtt{D}_{i}$ can be immediately recovered
from the eigenvalue decomposition of $\mathbf{Q}_{\mathtt{D}_{i}}^{\star}$
\cite{Boyd:ConvexOpt:2004} . However, this may not be the case since
the rank-1 constraints are dropped. Thus, a method to extract the
beamformer is required if a high-rank solution is obtained. For this
purpose, we adopt the randomization technique presented in \cite{Sidiropoulos:MulticastBeamforming:2006}
which is mentioned in line \ref{randomization} of Algorithm \ref{algo:iterativeMAXDET}
and briefly described as follows. We first generate a random (column)
vector $\mathbf{v}_{\mathtt{D}_{i}}$ whose elements are independently
and uniformly distributed on the unit circle in the complex plane,
and then calculate the eigen-decomposition of $\mathbf{Q}_{\mathtt{D}_{i}}^{\star}$
as $\mathbf{Q}_{\mathtt{D}_{i}}^{\star}=\mathbf{U}_{\mathtt{D}_{i}}\mathbf{\Sigma}_{\mathtt{D}_{i}}\mathbf{U}_{\mathtt{D}_{i}}^{H}$.
Next a beamformer is taken as $\mathbf{\tilde{w}}_{\mathtt{D}_{i}}=\mathbf{U}_{\mathtt{D}_{i}}\mathbf{\Sigma}_{\mathtt{D}_{i}}^{\nicefrac{1}{2}}\mathbf{v}_{\mathtt{D}_{i}}$,
which is feasible to the original design problem since $||\mathbf{\tilde{w}}_{\mathtt{D}_{i}}||_{2}^{2}=\tr(\mathbf{U}_{\mathtt{D}_{i}}\mathbf{\Sigma}_{\mathtt{D}_{i}}^{\nicefrac{1}{2}}\mathbf{v}_{\mathtt{D}_{i}}\mathbf{v}_{\mathtt{D}_{i}}^{H}\mathbf{\Sigma}_{\mathtt{D}_{i}}^{\nicefrac{1}{2}}\mathbf{U}_{\mathtt{D}_{i}}^{H})=\tr(\mathbf{Q}_{\mathtt{D}_{i}}^{\star})$
\cite{Sidiropoulos:MulticastBeamforming:2006}. The obtained beamformer
$\mathbf{\tilde{w}}_{\mathtt{D}_{i}}$ is then used to compute the
resulting sum rate. We repeat this process for a number of randomization
samples and pick up the one that offers the best sum rate. Our numerical
results have shown that the high-rank solutions of $\{\mathbf{Q}_{\mathtt{D}_{i}}\}$
only occur when $\sigma_{\mathtt{SI}}^{2}$ is sufficiently large,
which is not of practical importance since this is not the interesting
case for the full-duplex systems. When $\rank(\mathbf{Q}_{\mathtt{D}_{i}})>1$,
we also obverse that the largest eigenvalue significantly dominates
the remaining ones. More explicitly, the largest eigenvalue is always
$10$ times larger then the second largest one, meaning that $\mathbf{Q}_{\mathtt{D}_{i}}$
is not far from a rank-1 matrix. This explains the fact that the beamforming
vectors returned by the randomization method offer a performance very
close to that of the relaxed problem. Explicitly, the extracted solutions
achieve a spectral efficiency performance always higher than $95\%$
of the upper bound given by the relaxed problem.

Although the objective in \eqref{eq:MAXDET} is not a linear function
with respect to the design parameters as in the original work of \cite{FrankWolfe:1956},
\eqref{eq:MAXDET} can be equivalently transformed into the problem
of maximizing an affine function over a convex set as $\underset{\omega,\mathbf{Q},\mathbf{q}}{\max}\{\omega-g^{(n)}(\mathbf{Q},\mathbf{q})|h(\mathbf{Q},\mathbf{q})\geq\omega,\eqref{eq:SE:c1},\eqref{eq:SE:c2},\eqref{eq:SE:c3}\}$.
Thus, Algorithm \ref{algo:iterativeMAXDET} can be considered as a
variant of the FW method. It has been reported in many studies that
the type of FW methods can efficiently exploit the hidden convexity
of the problem \cite{Chris:LinearPrecoding:2010,Kha:PowerAllocation:DCprogramming:2012,Nam:SZF:2012}.
Thus, the same results as the FW-type method can also be expected
in the first proposed design algorithm. However, solvers for MAXDET
programs are quite limited, compared to their counterparts for SDPs.%
\footnote{The dedicated solver for the MAXDET problem in \eqref{eq:MAXDET}
is SDPT3 \cite{sdpt3}. In fact, CVX solves this type of problems
using a succesive convex approximate method, allowing us to choose
other SDP solvers, e.g., \cite{sedumi}. However, this method can
be slow and is still in an experimental stage.%
} Because none of the general convex program solvers are perfect for
all problems, a more flexible choice of a problem solver is of practical
importance.

\subsection{\label{sub:SDP-Approximation} Iterative SDP-based algorithm}

Motivated by the discussion above, we propose in this subsection an
iterative SDP-based algorithm to solve the relaxed problem of \eqref{eq:Original:SE}.
Specifically, based on the general framework of the SPCA method and
proper transformations, we can iteratively approximate the relaxed
problem of \eqref{eq:Original:SE} by an SDP in each iteration. The
second approach allows us to take advantage of a wide class of SDP
solvers which are more and more efficient due to continuing progress
in semidefinite programming. To begin with, due to the monotonicity
of the $\log$ function, we first reformulate the relaxed problem
of \eqref{eq:Original:SE} as
\begin{equation}
\begin{array}{rl}
\underset{\{\mathbf{Q}_{\mathtt{D}_{i}}\},\{q_{\mathtt{U}_{j}}\}}{\maximize} & \prod_{i=1}^{K_{\mathtt{D}}}(1+\gamma_{\mathtt{D}_{i}})\prod_{j=1}^{K_{\mathtt{U}}}(1+\gamma_{\mathtt{U}_{j}})\\
\st & \eqref{eq:SE:c1},\;\eqref{eq:SE:c2},\;\eqref{eq:SE:c3}
\end{array}\label{eq:Problem:Monotomic}
\end{equation}
which then can be rewritten as \begin{IEEEeqnarray}{cl}\IEEEyesnumber\label{eq:Problem:Epi}\underset{{\scriptstyle \{\mathbf{Q}_{\mathtt{D}_{i}}\},\{q_{\mathtt{U}_{j}}\},\{t_{\mathtt{D}_{i}}\},\{t_{\mathtt{U}_{j}}\}}}{\maximize} &  \quad \prod_{i=1}^{K_{\mathtt{D}}}t_{\mathtt{D}_{i}}\prod_{j=1}^{K_{\mathtt{U}}}t_{\mathtt{U}_{j}}\IEEEeqnarraynumspace\IEEEeqnarraynumspace\IEEEyessubnumber\label{eq:Pro:Epi:Ojb}\\ \st &  \quad 1+\gamma_{\mathtt{D}_{i}}\geq t_{\mathtt{D}_{i}},\; i=1,\ldots,K_{\mathtt{D}},\IEEEeqnarraynumspace\IEEEyessubnumber\label{eq:pro:epi:c1}\\  &  \quad 1+\gamma_{\mathtt{U}_{j}}\geq t_{\mathtt{U}_{j}},\; j=1,\ldots,K_{\mathtt{U}},\IEEEeqnarraynumspace\IEEEyessubnumber\label{eq:pro:epi:c2}\\  &  \quad t_{\mathtt{D}_{i}}\geq1,\forall i;t_{\mathtt{U}_{j}}\geq1,\forall j,\IEEEeqnarraynumspace\IEEEyessubnumber\label{eq:sinr:reg}\\  &  \quad \eqref{eq:SE:c1},\;\eqref{eq:SE:c2},\;\eqref{eq:SE:c3} \IEEEyessubnumber\end{IEEEeqnarray}
by using the epigraph form of \eqref{eq:Problem:Monotomic} \cite{Boyd:ConvexOpt:2004}.
Note that maximizing a product of variables admits an SOC representation
\cite{Nam:WSRM:SOCP:2012,Lobo:SOCP:1998}. Thus, we only need to deal
with the nonconvex constraints in \eqref{eq:pro:epi:c1} and \eqref{eq:pro:epi:c2}.
Let us treat the constraint \eqref{eq:pro:epi:c1} first. It is without
loss of optimality to replace \eqref{eq:pro:epi:c1} by following
two constraints\begin{subequations}\label{eq:split:c1}
\begin{eqnarray}
\sigma_{n}^{2}+\sum_{k=1}^{K_{\mathtt{D}}}\mathbf{h}_{\mathtt{D}_{i}}^{H}\mathbf{Q}_{\mathtt{D}_{k}}\mathbf{h}_{\mathtt{D}_{i}}+\sum_{j=1}^{K_{\mathtt{U}}}q_{\mathtt{U}_{j}}|g_{ji}|^{2} & \geq & t_{\mathtt{D}_{i}}\beta_{\mathtt{D}_{i}},\label{eq:pro:epi:c11}\\
\sigma_{n}^{2}+\sum_{k\neq i}^{K_{\mathtt{D}}}\mathbf{h}_{\mathtt{D}_{i}}^{H}\mathbf{Q}_{\mathtt{D}_{k}}\mathbf{h}_{\mathtt{D}_{i}}+\sum_{j=1}^{K_{\mathtt{U}}}q_{\mathtt{U}_{j}}|g_{ji}|^{2} & \leq & \beta_{\mathtt{D}_{i}}\label{eq:pro:epi:c12}
\end{eqnarray}
\end{subequations} where $\beta_{\mathtt{D}_{i}}$ is newly introduced
variable and can be considered as the soft interference threshold
of $\mathtt{D}_{i}$. The equivalence between \eqref{eq:pro:epi:c1}
and the two inequalities in \eqref{eq:pro:epi:c11} and \eqref{eq:pro:epi:c12}
follows the same arguments as in \cite{Nam:WSRM:SOCP:2012} which
can be justified as follows. At optimum, suppose the constraint in
\eqref{eq:pro:epi:c12} holds with inequality, i.e., $\sigma_{n}^{2}+\sum_{k\neq i}^{K_{\mathtt{D}}}\mathbf{h}_{\mathtt{D}_{i}}^{H}\mathbf{Q}_{\mathtt{D}_{k}}\mathbf{h}_{\mathtt{D}_{i}}+\sum_{j=1}^{K_{\mathtt{U}}}q_{\mathtt{U}_{j}}|g_{ji}|^{2}<\beta_{\mathtt{D}_{i}}$.
Then, we form a new pair $(\bar{\beta}_{\mathtt{D}_{i}},\bar{t}_{\mathtt{D}_{i}})$
as $\bar{\beta}_{\mathtt{D}_{i}}\triangleq\beta_{\mathtt{D}_{i}}/c$
and $\bar{t}_{\mathtt{D}_{i}}\triangleq ct_{\mathtt{D}_{i}}$ where
$c$ is a positive constant. Obviously, there exists a given $c>1$
such that \eqref{eq:pro:epi:c12} is still met when $\beta_{\mathtt{D}_{i}}$
is replaced by $\bar{\beta}_{\mathtt{D}_{i}}$. Since $\bar{\beta}_{\mathtt{D}_{i}}\bar{t}_{\mathtt{D}_{i}}=\beta_{\mathtt{D}_{i}}t_{\mathtt{D}_{i}}$,
i.e., the right side of \eqref{eq:pro:epi:c11} remains the same,
the constraint in \eqref{eq:pro:epi:c11} is still satisfied. However,
since $\bar{t}_{\mathtt{D}_{i}}>t_{\mathtt{D}_{i}}$ with $c>1$,
a strictly higher objective of the design problem is obtained. This
contradicts with assumption that we already obtained the optimal objective.
Likewise, we can decompose \eqref{eq:pro:epi:c2} into \begin{subequations}\label{eq:split:c2}
\begin{eqnarray}
x_{\mathtt{U}_{j}}^{2}\mathbf{h}_{\mathtt{U}_{j}}^{H}\mathbf{X}_{\mathtt{U}_{j}}^{-1}\mathbf{h}_{\mathtt{U}_{j}} & \geq & t_{\mathtt{U}_{j}}-1,\label{eq:pro:epi:c21}\\
q_{\mathtt{U}_{j}} & \geq & x_{\mathtt{U}_{j}}^{2}\label{eq:pro:epi:c22}
\end{eqnarray}
\end{subequations} where $\mathbf{X}_{\mathtt{U}_{j}}\triangleq\sigma_{n}^{2}\mathbf{I}+\sum_{m>j}^{K_{\mathtt{U}}}q_{\mathtt{U}_{m}}\mathbf{h}_{\mathtt{U}_{m}}\mathbf{h}_{\mathtt{U}_{m}}^{H}+\sum_{i=1}^{K_{\mathtt{D}}}\mathbf{H}_{\mathtt{SI}}\mathbf{Q}_{\mathtt{D}_{i}}\mathbf{H}_{\mathtt{SI}}^{H}$
and $x_{\mathtt{U}_{j}}$ is an auxiliary variable. The purpose of
introducing slack variable $x_{\mathtt{U}_{j}}$ will be clear shortly
when we show that it is necessary to arrive at an SDP formulation.
Now, we can equivalently transform \eqref{eq:Problem:Epi} into a
more tractable form as\begin{IEEEeqnarray}{cl}\IEEEyesnumber\label{eq:SE:reform}\underset{\underset{\mathbf{\beta}_{\mathtt{D}},\mathbf{x}_{\mathtt{U}}}{{\scriptstyle \mathbf{Q},\mathbf{q},\mathbf{t}_{\mathtt{D}},\mathbf{t}_{\mathtt{U}},}}}{\maximize} &  \quad \prod_{i=1}^{K_{\mathtt{D}}}t_{\mathtt{D}_{i}}\prod_{j=1}^{K_{\mathtt{U}}}t_{\mathtt{U}_{j}}\IEEEyessubnumber\label{eq:SE:reform:obj}\\ \st &  \quad \sigma_{n}^{2}+\sum_{k=1}^{K_{\mathtt{D}}}\mathbf{h}_{\mathtt{D}_{i}}^{H}\mathbf{Q}_{\mathtt{D}_{k}}\mathbf{h}_{\mathtt{D}_{i}}+\sum_{j=1}^{K_{\mathtt{U}}}q_{\mathtt{U}_{j}}\bigl|g_{ji}\bigr|^{2}\nonumber \\  &  \quad \;\quad\geq f(t_{\mathtt{D}_{i}},\beta_{\mathtt{D}_{i}}),\;\forall i=1,\ldots,K_{\mathtt{D}},\IEEEyessubnumber\label{eq:SE:reform:c1}\\  &  \quad \sigma_{n}^{2}+\sum_{k\neq i}^{K_{\mathtt{D}}}\mathbf{h}_{\mathtt{D}_{i}}^{H}\mathbf{Q}_{\mathtt{D}_{k}}\mathbf{h}_{\mathtt{D}_{i}}+\sum_{j=1}^{K_{\mathtt{U}}}q_{\mathtt{U}_{j}}\bigl|g_{ji}\bigr|^{2}\nonumber \\  &  \quad \;\quad\leq\beta_{\mathtt{D}_{i}},\;\forall i=1,\ldots,K_{\mathtt{D}},\IEEEyessubnumber\label{eq:SE:reform:c2}\\  &  \quad g(x_{\mathtt{U}_{j}}^{2},\mathbf{Q},\mathbf{q})\geq t_{\mathtt{U}_{j}}-1,\;\forall j=1,\ldots,K_{\mathtt{U}},\IEEEeqnarraynumspace\IEEEyessubnumber\label{eq:SE:reform:c3}\\  &  \quad q_{\mathtt{U}_{j}}\geq x_{\mathtt{U}_{j}}^{2},\;\forall j=1,\ldots,K_{\mathtt{U}},\IEEEyessubnumber\label{eq:SE:reform:c4}\\  &  \quad \eqref{eq:SE:c1},\;\eqref{eq:SE:c2},\;\eqref{eq:SE:c3},\;\eqref{eq:sinr:reg}\IEEEyessubnumber\end{IEEEeqnarray}
where $f(t_{\mathtt{D}_{i}},\beta_{\mathtt{D}_{i}})\triangleq t_{\mathtt{D}_{i}}\beta_{\mathtt{D}_{i}}$,
$g(x_{\mathtt{U}_{j}}^{2},\mathbf{Q},\mathbf{q})\triangleq x_{\mathtt{U}_{j}}^{2}\mathbf{h}_{\mathtt{U}_{j}}^{H}\mathbf{X}_{\mathtt{U}_{j}}^{-1}\mathbf{h}_{\mathtt{U}_{j}}$,
and $\mathbf{Q}$, $\mathbf{q}$, $\mathbf{t}_{\mathtt{D}}$, $\mathbf{t}_{\mathtt{U}}$,
$\mathbf{\beta}_{\mathtt{D}}$, $\mathbf{x}_{\mathtt{U}}$ are the
symbolic notations that denote the sets of optimization variables
$\{\mathbf{Q}_{\mathtt{D}_{i}}\}$, $\{q_{\mathtt{U}_{j}}\}$, $\{t_{\mathtt{D}_{i}}\}$,$\{t_{\mathtt{U}_{j}}\}$,
$\{\beta_{\mathtt{D}_{i}}\}$, $\{x_{\mathtt{U}_{j}}\}$, respectively.

We note that the constraints in \eqref{eq:SE:reform:c2} and \eqref{eq:SE:reform:c4}
are linear and SOC ones, respectively. Consequently, the barrier to
solving \eqref{eq:SE:reform} is due to the nonconvexity in \eqref{eq:SE:reform:c1}
and \eqref{eq:SE:reform:c3}. In what follows, we will present a low-complexity
approach that locally solves \eqref{eq:SE:reform}. Toward this end
we resort to an iterative algorithm based on SPCA. To show this, let
us tackle the nonconvex constraint \eqref{eq:SE:reform:c1} first.
Note that $f(t_{\mathtt{D}_{i}},\beta_{\mathtt{D}_{i}})$ is neither
a convex nor concave function of $t_{\mathtt{D}_{i}}$ and $\beta_{\mathtt{D}_{i}}$.
Fortunately, in the spirit of \cite{Nam:WSRM:SOCP:2012,Beck:SCA:2010},
we recall the following inequality
\begin{equation}
f(t_{\mathtt{D}_{i}},\beta_{\mathtt{D}_{i}})\leq F(t_{\mathtt{D}_{i}},\beta_{\mathtt{D}_{i}},\psi_{\mathtt{D}_{i}}^{(n)})=\frac{1}{2\psi_{\mathtt{D}_{i}}^{(n)}}t_{\mathtt{D}_{i}}^{2}+\frac{\psi_{\mathtt{D}_{i}}^{(n)}}{2}\beta_{\mathtt{D}_{i}}^{2}\label{eq:c1:CVX:UB}
\end{equation}
which holds for every $\psi_{\mathtt{D}_{i}}^{(n)}>0$. The right
side of \eqref{eq:c1:CVX:UB} is called a convex upper estimate of
$f(t_{\mathtt{D}_{i}},\beta_{\mathtt{D}_{i}})$. The approximation
shown in \eqref{eq:c1:CVX:UB} deserves some comments. First, it is
straightforward to note that $f(t_{\mathtt{D}_{i}},\beta_{\mathtt{D}_{i}})=F(t_{\mathtt{D}_{i}},\beta_{\mathtt{D}_{i}},\psi_{\mathtt{D}_{i}}^{(n)})$
when $\psi_{\mathtt{D}_{i}}^{(n)}=t_{\mathtt{D}_{i}}/\beta_{\mathtt{D}_{i}}$.%
\footnote{Since $t_{\mathtt{D}_{i}}\geq1$ and $\beta_{\mathtt{D}_{i}}\geq\sigma_{n}^{2}>0$
(from \eqref{eq:pro:epi:c12}) and both of them are bounded above
(i.e., $t_{\mathtt{D}_{i}}<+\infty$ and $\beta_{\mathtt{D}_{i}}<+\infty$)
due to the transmit power constraint at the BS, the value of $\psi_{\mathtt{D}_{i}}^{(n)}$
is well defined. %
} Moreover, with this selection of $\psi_{\mathtt{D}_{i}}^{(n)}$,
one can easily check that the first derivative of $F(t_{\mathtt{D}_{i}},\beta_{\mathtt{D}_{i}},\psi_{\mathtt{D}_{i}}^{(n)})$
with respect to $t_{\mathtt{D}_{i}}$ or $\beta_{\mathtt{D}_{i}}$
is equal to that of $f(t_{\mathtt{D}_{i}},\beta_{\mathtt{D}_{i}})$,
i.e., $\nabla F(t_{\mathtt{D}_{i}},\beta_{\mathtt{D}_{i}},\psi_{\mathtt{D}_{i}}^{(n)})=\nabla f(t_{\mathtt{D}_{i}},\beta_{\mathtt{D}_{i}})$.
These two properties are important to establish the local convergence
of the second iterative algorithm which is deferred to the Appendix.

Now we turn our attention to \eqref{eq:SE:reform:c3}, which is equivalent
to $t_{\mathtt{U}_{j}}-1-g(x_{\mathtt{U}_{j}}^{2},\mathbf{Q},\mathbf{q})\leq0$.
First, we note that $g(x_{\mathtt{U}_{j}}^{2},\mathbf{Q},\mathbf{q})$
is jointly convex in the involved variables. As proof, consider the
epigraph of $g(x_{\mathtt{U}_{j}}^{2},\mathbf{Q},\mathbf{q})$ which
is given by \cite{Boyd:ConvexOpt:2004}
\begin{equation}
\bigl\{\bigl(\alpha,x_{\mathtt{U}_{j}}^{2},\mathbf{Q},\mathbf{q}\bigr)|\alpha\geq x_{\mathtt{U}_{j}}^{2}\mathbf{h}_{\mathtt{U}_{j}}^{H}\mathbf{X}_{\mathtt{U}_{j}}^{-1}\mathbf{h}_{\mathtt{U}_{j}}\bigr\}.\label{eq:epi}
\end{equation}
By Schur complement \cite{Dattorro:ConvexOpt:2011}, \eqref{eq:epi}
is equivalent to
\begin{multline}
\begin{bmatrix}\alpha & x_{\mathtt{U}_{j}}\mathbf{h}_{\mathtt{U}_{j}}^{H}\\
x_{\mathtt{U}_{j}}\mathbf{h}_{\mathtt{U}_{j}} & \mathbf{X}_{\mathtt{U}_{j}}
\end{bmatrix}=\\
\begin{bmatrix}\alpha & x_{\mathtt{U}_{j}}\mathbf{h}_{\mathtt{U}_{j}}^{H}\\
x_{\mathtt{U}_{j}}\mathbf{h}_{\mathtt{U}_{j}} & \sigma_{n}^{2}\mathbf{I}+{\displaystyle \sum_{m>j}^{K_{\mathtt{U}}}}q_{\mathtt{U}_{m}}\mathbf{h}_{\mathtt{U}_{m}}\mathbf{h}_{\mathtt{U}_{m}}^{H}+{\displaystyle \sum_{i=1}^{K_{\mathtt{D}}}}\mathbf{H}_{\mathtt{SI}}\mathbf{Q}_{\mathtt{D}_{i}}\mathbf{H}_{\mathtt{SI}}^{H}
\end{bmatrix}\succeq\mathbf{0}.
\end{multline}
 Since the epigraph of $g(x_{\mathtt{U}_{j}}^{2},\mathbf{Q},\mathbf{q})$
is representable by linear matrix inequality which is a convex set,
so is $g(x_{\mathtt{U}_{j}}^{2},\mathbf{Q},\mathbf{q})$ \cite{Boyd:ConvexOpt:2004}.
Now a convex upper bound of the term $-g(x_{\mathtt{U}_{j}}^{2},\mathbf{Q},\mathbf{q})$
in \eqref{eq:SE:reform:c3} can be found as its first order approximation
at a neighborhood of $(x_{\mathtt{U}_{j}}^{(n)},\mathbf{Q}^{(n)},\mathbf{q}^{(n)})$,
i.e.,
\begin{multline}
-g(x_{\mathtt{U}_{j}}^{2},\mathbf{Q},\mathbf{q})\leq G\Bigl(x_{\mathtt{U}_{j}},\mathbf{Q},\mathbf{q},x_{\mathtt{U}_{j}}^{(n)},\mathbf{Q}^{(n)},\mathbf{q}^{(n)}\Bigr)\\
=-\biggl\{ g(x_{\mathtt{U}_{j}}^{(n)},\mathbf{Q}^{(n)},\mathbf{q}^{(n)})+2x_{\mathtt{U}_{j}}^{(n)}\mathbf{h}_{\mathtt{U}_{j}}^{H}(\mathbf{X}_{\mathtt{U}_{j}}^{(n)})^{-1}\mathbf{h}_{\mathtt{U}_{j}}\bigl(x_{\mathtt{U}_{j}}-x_{\mathtt{U}_{j}}^{(n)}\bigr)\\
-\tr\Bigl[\Bigl((x_{\mathtt{U}_{j}}^{(n)})^{2}\bigl(\mathbf{X}_{\mathtt{U}_{j}}^{(n)}\bigr)^{-1}\mathbf{h}_{\mathtt{U}_{j}}\mathbf{h}_{\mathtt{U}_{j}}^{H}\bigl(\mathbf{X}_{\mathtt{U}_{j}}^{(n)}\bigr)^{-1}\Bigr)\Bigl(\mathbf{X}_{\mathtt{U}_{j}}-\mathbf{X}_{\mathtt{U}_{j}}^{(n)}\Bigr)\Bigr]\biggr\}\label{eq:c3:CVX:UB}
\end{multline}
where $\mathbf{X}_{\mathtt{U}_{j}}$ is replaced by the affine function
of $\mathbf{Q}$ and $\mathbf{q}$ defined below \eqref{eq:split:c2}
and we have used the fact that $\nabla_{\mathbf{A}}\,\mathbf{a}^{H}\mathbf{A}^{-1}\mathbf{b}=-\mathbf{A}^{-1}\mathbf{a}\mathbf{b}^{H}\mathbf{A}^{-1}$
for $\mathbf{A}\succeq\mathbf{0}$ \cite{Dattorro:ConvexOpt:2011}.

\begin{algorithm}[tb]
\caption{Iterative SDP-based algorithm.}
\label{algo:IterativeSDP}\begin{algorithmic}[1]
\renewcommand{\algorithmicrequire}{\textbf{Initialization:}} \REQUIRE
\STATE Generate initial points for $\psi_{\mathtt{D}_{i}}^{(0)}$
and $\mathbf{Q}_{\mathtt{D}_{i}}^{(0)}$ for $i=1,\ldots,K_{\mathtt{D}}$;
and $q_{\mathtt{U}_{j}}^{(0)}$ and $x_{\mathtt{U}_{j}}^{(0)}$ for
$j=1,\ldots,K_{\mathtt{U}}$.
\STATE Set $n:=0$.
\renewcommand{\algorithmicrequire}{\textbf{Iterative procedure:}}
\REQUIRE
\REPEAT
\STATE Solve \eqref{eq:SDP} to find optimal solutions $\mathbf{Q}_{\mathtt{D}_{i}}^{\star}$,
$t_{\mathtt{D}_{i}}^{\star}$, and $\beta_{\mathtt{D}_{i}}^{\star}$
for $i=1,\ldots,K_{\mathtt{D}}$, and $q_{\mathtt{U}_{j}}^{\star}$,
and $x_{\mathtt{U}_{j}}^{\star}$ for $j=1,\ldots,K_{\mathtt{U}}$.
\STATE Set $n:=n+1$.
\STATE Update : $\psi_{\mathtt{D}_{i}}^{(n)}:=t_{\mathtt{D}_{i}}^{\star}/\beta_{\mathtt{D}_{i}}^{\star}$;
$x_{\mathtt{U}_{j}}^{(n)}:=x_{\mathtt{U}_{j}}^{\star}$; $\mathbf{Q}_{\mathtt{D}_{i}}^{(n)}:=\mathbf{Q}_{\mathtt{D}_{i}}^{\star}$;
$q_{\mathtt{U}_{j}}^{(n)}:=q_{\mathtt{U}_{j}}^{\star}$.
\UNTIL Convergence.
\renewcommand{\algorithmicrequire}{\textbf{Finalization:}} \REQUIRE
\STATE Perform randomization to extract a rank-1 solution as in Algorithm
\ref{algo:iterativeMAXDET}.
\end{algorithmic}
\end{algorithm}

The mathematical discussions above imply that the convex approximate
problem at iteration $n+1$ of the second iterative design approach
is the following\begin{IEEEeqnarray}{cl}\IEEEyesnumber\label{eq:SDP}\underset{\underset{\mathbf{\beta}_{\mathtt{D}},\mathbf{x}_{\mathtt{U}}}{{\scriptstyle \mathbf{Q},\mathbf{q},\mathbf{t}_{\mathtt{D}},\mathbf{t}_{\mathtt{U}},}}}{\maximize} &  \quad \prod_{i=1}^{K_{\mathtt{D}}}t_{\mathtt{D}_{i}}\prod_{j=1}^{K_{\mathtt{U}}}t_{\mathtt{U}_{j}}\IEEEyessubnumber\label{eq:Obj:Product_SOCconstr}\\ \st &  \quad F(t_{\mathtt{D}_{i}},\beta_{\mathtt{D}_{i}},\psi_{\mathtt{D}_{i}}^{(n)})\leq\sigma_{n}^{2}+\sum_{k=1}^{K_{\mathtt{D}}}\mathbf{h}_{\mathtt{D}_{i}}^{H}\mathbf{Q}_{\mathtt{D}_{k}}\mathbf{h}_{\mathtt{D}_{i}}\nonumber \\  &  \quad \;+\sum_{j=1}^{K_{\mathtt{U}}}q_{\mathtt{U}_{j}}\bigl|g_{ji}\bigr|^{2},\forall i=1,\ldots,K_{\mathtt{D}},\IEEEyessubnumber\label{eq:SDP_1}\\  &  \quad G\bigl(x_{\mathtt{U}_{j}},\mathbf{Q},\mathbf{q},x_{\mathtt{U}_{j}}^{(n)},\mathbf{Q}^{(n)},\mathbf{q}^{(n)}\bigr)\nonumber \\  &  \quad \;\leq1-t_{\mathtt{U}_{j}},\forall j=1,\ldots,K_{\mathtt{U}},\IEEEyessubnumber\label{eq:SDP_3}\\  &  \quad \sigma_{n}^{2}+\sum_{k\neq i}^{K_{\mathtt{D}}}\mathbf{h}_{\mathtt{D}_{i}}^{H}\mathbf{Q}_{\mathtt{D}_{k}}\mathbf{h}_{\mathtt{D}_{i}}\nonumber\\  &  \quad \;+\sum_{j=1}^{K_{\mathtt{U}}}q_{\mathtt{U}_{j}}\bigl|g_{ji}\bigr|^{2}\leq\beta_{\mathtt{D}_{i}},\;\forall i=1,\ldots,K_{\mathtt{D}},\IEEEeqnarraynumspace\IEEEyessubnumber\label{eq:SDP_4}\\  &  \quad q_{\mathtt{U}_{j}}\geq x_{\mathtt{U}_{j}}^{2},\;\forall j=1,\ldots,K_{\mathtt{U}},\IEEEyessubnumber\label{eq:SDP_5}\\  &  \quad 0\leq q_{\mathtt{U}_{j}}\leq\overline{q}_{\mathtt{U}_{j}},\forall j=1,\ldots,K_{\mathtt{U}},\IEEEyessubnumber\label{eq:SDP_6}\\  &  \quad \sum_{i=1}^{K_{\mathtt{D}}}\tr(\mathbf{Q}_{\mathtt{D}_{i}})\leq P_{\mathtt{BS}},\IEEEyessubnumber\label{eq:SDP_7}\\  &  \quad \mathbf{Q}_{\mathtt{D}_{i}}\succeq0,\forall i=1,\ldots,K_{\mathtt{D}},\IEEEyessubnumber\label{eq:SDP_8}\\  &  \quad \hspace{-35pt}t_{\mathtt{D}_{i}}\geq1,\forall i=1,\ldots,K_{\mathtt{D}};t_{\mathtt{U}_{j}}\geq1,\forall j=1,\ldots,K_{\mathtt{U}}\IEEEyessubnumber.\end{IEEEeqnarray}

After the iterative procedure terminates, the randomization trick
may be applied to extract a rank-1 solution as in Algorithm \ref{algo:iterativeMAXDET}.
The proposed iterative SDP-based algorithm is summarized in Algorithm
\ref{algo:IterativeSDP}.

The convergence results of Algorithms \ref{algo:iterativeMAXDET}
and \ref{algo:IterativeSDP} are stated in the following theorem whose
proof is given in the Appendix.
\begin{thm}
\label{thm:converge}Algorithms \ref{algo:iterativeMAXDET} and \ref{algo:IterativeSDP}
produce a sequence of solutions converging to a KKT point of \eqref{eq:DCform}
and \eqref{eq:Problem:Epi}, respectively.
\end{thm}
As mentioned in \cite{Beck:SCA:2010}, the SPCA method can start with
an infeasible initial point. However, it is desired to generate initial
values for $\mathbf{Q}_{\mathtt{D}_{i}}^{(0)}$, $q_{\mathtt{U}_{j}}^{(0)}$
, $\psi_{\mathtt{D}_{i}}^{(0)}$ and $x_{\mathtt{U}_{j}}^{(0)}$ such
that Algorithm \ref{algo:IterativeSDP} is guaranteed to be solvable
in the first iteration. For this purpose, we first randomly generate
$\mathbf{Q}_{\mathtt{D}_{i}}^{(0)}\succeq0$ for $i=1,\ldots,K_{\mathtt{D}}$
and $q_{\mathtt{U}_{j}}^{(0)}$ in the range from $0$ to $\overline{q}_{\mathtt{U}_{j}}$
for $j=1,\ldots,K_{\mathtt{U}}$. If necessary, $\mathbf{Q}_{\mathtt{D}_{i}}^{(0)}$
is scaled so that the constraint \eqref{eq:SDP_7} is satisfied. Then,
$x_{\mathtt{U}_{j}}^{(0)}$ is calculated as $x_{\mathtt{U}_{j}}^{(0)}=\sqrt{q_{\mathtt{U}_{j}}^{(0)}}$
and $\psi_{\mathtt{D}_{i}}^{(0)}$ is set to $t_{\mathtt{D}_{i}}^{(0)}/\beta_{\mathtt{D}_{i}}^{(0)}$
where $t_{\mathtt{D}_{i}}^{(0)}$ and $\beta_{\mathtt{D}_{i}}^{(0)}$
are computed from \eqref{eq:pro:epi:c1} and \eqref{eq:SDP_4} by
setting the inequalities to equalities, respectively.

At the first look, the SDP solved at each iteration in Algorithm \ref{algo:IterativeSDP}
has more optimization variables due to some slack variables introduced.
Thus, the theoretical (worst case) computational complexity of Algorithm
\ref{algo:IterativeSDP} could possibly be higher than that of Algorithm
\ref{algo:iterativeMAXDET}. We note that the complexity of the two
proposed methods mainly depends on the semidefinite constraints $\mathbf{Q}_{\mathtt{D}_{i}}\succeq0$,
$\forall i=1,\ldots,K_{\mathtt{D}}$. That is to say, the per iteration
complexity formulation used in Algorithm \ref{algo:IterativeSDP}
just slightly requires higher complexity than Algorithm \ref{algo:iterativeMAXDET}.
As aforementioned, the advantage of the second proposed algorithm
is that it allows us to make use of efficient SDP solvers such as
SEDUMI and MOSEK. Alternatively, we can use both proposed two approaches
in parallel for solving the original problem. The solving process
can be terminated if one of the algorithms has converged. It is also
possible to solve the problem until both methods converge and choose
the better solution. More insights on the computational complexity
of the iterative MAXDET- and SDP- based algorithms are given in Section
\ref{sec:numerical results}.

In closing this section two remarks are in order. First, the proposed
algorithms are also valid for macro cell full-duplex systems (if practically
implementable). Our emphasis on small cell setups is merely due to
current practical limitations. Second, the mathematical presentation
can be slightly modified to arrive at a centralized joint beamformer
design for a multicell deployment scenario. Specifically, if all the
CSI can be timely forwarded to the centralized processing unit, a
joint design is straightforward. Obviously, distributed solutions
are more interesting from a practical perspective and will be explored
in the follow-up work. \vspace{0pt}

\section{\label{sec:numerical results}Numerical Results }

\subsection{\label{sub:Convergence-and-Complexity}Convergence and Complexity
Comparison}

In the first experiment we compare the complexity and the convergence
rate of Algorithms \ref{algo:iterativeMAXDET} and \ref{algo:IterativeSDP}
proposed in Section \ref{sec:designs} for two cases, the first case
for independent and identically distributed (i.i.d) channel model
and the second case for realistic channel model generated in Section
\ref{sub:Spectral-Efficiency-Performance}. In the first case, each
entry of the channel vectors $\mathbf{h}_{\mathtt{D}_{i}}$, $\mathbf{h}_{\mathtt{U}_{i}}$,
and $g_{ji}$ follows the i.i.d zero mean and unit variance Gaussian
distribution. The noise power is taken as $\sigma_{n}^{2}=1$ and
the maximum transmit power at the BS and uplink users are set to $P_{\mathtt{BS}}=q_{\mathtt{U}_{j}}=20$
dBW for all $\mathtt{U}_{j}$. This setting resembles the case where
the average signal to noise ratio (SNR) at transmitter sides is $20$
dB. In the second case, the specific parameters are taken from Table
\ref{table:sysparams} and the allowable transmit power at the BS
and the users in the uplink channel are fixed at $P_{\mathtt{BS}}=q_{\mathtt{U}_{j}}=10$
dBm.

An accurate model for the self-interference channel plays an important
role in evaluating the SE performance of full-duplex systems. Thus,
theoretical studies and practical measurements on this issue are of
significant importance and call for more research efforts. A pioneer
practical experiment on self-interference channel model has been carried
out in \cite{Duarte:Fullduplex:Experiment}. The main conclusion of
\cite{Duarte:Fullduplex:Experiment} is that the Rician probability
distribution with a small Rician factor should be used to characterize
the residual self-interference channel after self-interference cancellation
mechanisms. Hence, in this paper, $\mathbf{H}_{\mathtt{SI}}$ is generated
as $\mathcal{CN}_{N_{\mathtt{R}}N_{\mathtt{T}}}\Bigl(\sqrt{\frac{\sigma_{\mathtt{SI}}^{2}K}{1+K}}\bar{\mathbf{H}}_{\mathtt{SI}},\frac{\sigma_{\mathtt{SI}}^{2}}{1+K}I_{N_{\mathtt{R}}}\otimes I_{N_{\mathtt{T}}}\Bigr)$,
where $\otimes$ denotes the Kronecker product, $K$ is the Rician
factor, $\bar{\mathbf{H}}_{\mathtt{SI}}$ is a deterministic matrix,
and $\sigma_{\mathtt{SI}}^{2}$ is introduced to parameterize the
capability of a certain self-interference cancellation design.%
\footnote{Without loss of generality, we set $K=1$ and $\bar{\mathbf{H}}_{\mathtt{SI}}$
to be the matrix of all ones for all experiments. %
} In this model, $\sigma_{\mathtt{SI}}^{2}$ is the ratio of the average
self-interference power before and after the cancellation process
and its value is fixed at $-30$ dB for the first case and $-100$
dB for the second case in this numerical simulation.

\begin{figure}
\centering \subfigcapskip = 0cm \subfigure[Convergence rate for i.i.d channel realizations with $K_{\mathtt{D}}=K_{\mathtt{U}}=4$ and $N_{\mathtt{T}}=N_{\mathtt{R}}=4$. \label{fig:rateIVA}]{\includegraphics[width=1\columnwidth]{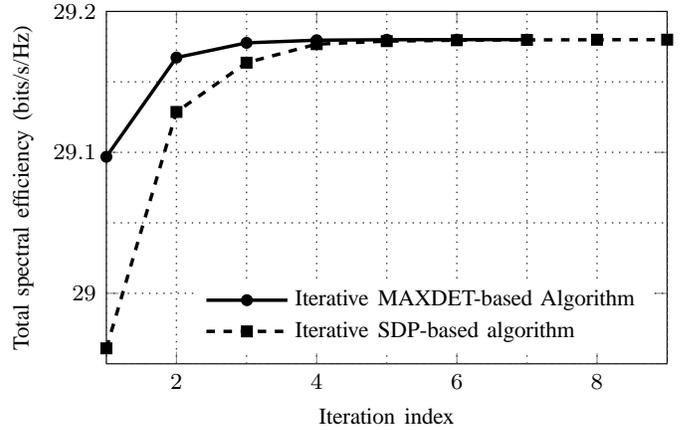}}	 \hspace{2mm}\subfigure[Convergence rate for channel realizations taken from the channel model in Section IV-B. In this setup, $N_{\mathtt{T}}=4$, $N_{\mathtt{R}}=2$, $K_{\mathtt{D}}=6$, and $K_{\mathtt{U}}=4$.  \label{fig:rateIVB}]{\includegraphics[width=1\columnwidth]{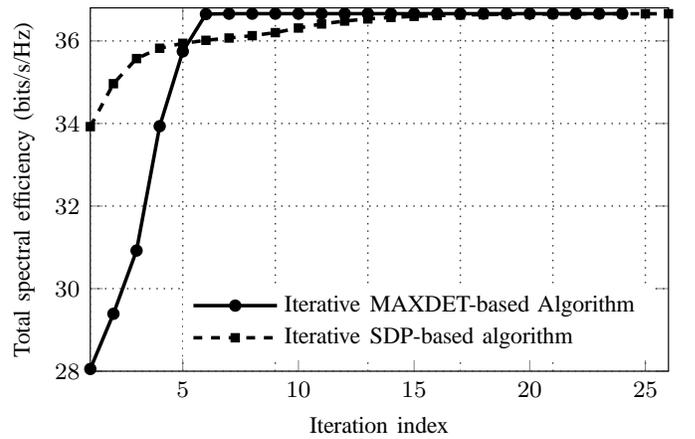}}\caption{\label{fig:convergence}Convergence rate of Algorithms \ref{algo:iterativeMAXDET}
and \ref{algo:IterativeSDP} for a set of random channel realizations. }
\end{figure}

Fig. \ref{fig:convergence} illustrates the convergence rate of Algorithms
\ref{algo:iterativeMAXDET} and \ref{algo:IterativeSDP} for a given
set of channel realizations generated randomly for the two cases.
Each point on the curves of Fig. \ref{fig:convergence} is obtained
by solving problems \eqref{eq:MAXDET} and \eqref{eq:SDP}, respectively.
The simulation settings are included in the figure caption for ease
of reference. Generally, we have observed that Algorithm \ref{algo:iterativeMAXDET}
requires fewer iterations to converge than Algorithm \ref{algo:IterativeSDP}.
This observation is probably attributed to the fact that Algorithm
\ref{algo:iterativeMAXDET} exploits the hidden convexity better since
it searches for an improved solution over the whole feasible set in
each iteration. We recall that SDPT3 is the dedicated solver for the
type of problems in \eqref{eq:MAXDET}, and thus the choice of optimization
software is limited for Algorithm \ref{algo:iterativeMAXDET}. A recent
work of \cite{Mittelmann:ConicSolver} has reported that, among common
general SDP solvers, SDPT3 is comparatively slow. The SDP formulation
in Algorithm \ref{algo:IterativeSDP} allows for use of faster SDP
solvers such as SeDuMi or MOSEK. In return, the total time of Algorithm
\ref{algo:IterativeSDP} to find a solution may be less than that
of Algorithm \ref{algo:iterativeMAXDET} which is illustrated in Table
\ref{tab:runtime}.

In Table \ref{tab:runtime}, we show the average run time (in seconds)
of Algorithms \ref{algo:iterativeMAXDET} and \ref{algo:IterativeSDP}
for the two channel models mentioned above. The stopping criterion
for the two algorithms is when the increase in the last $10$ iterations
is less than $10^{-5}$. All convex solvers considered in Table \ref{tab:runtime}
are set to their default values. We observe that the per iteration
solving time of Algorithm \ref{algo:IterativeSDP} is much less than
that of Algorithm \ref{algo:iterativeMAXDET}. Consequently, the total
solving time of Algorithm \ref{algo:IterativeSDP} is smaller than
that of Algorithm \ref{algo:iterativeMAXDET}, especially when used
with MOSEK solver.

\begin{table*}[t]
\caption{Average run time (in seconds) for i.i.d and realistic channel models
for various simulation setups. The proposed algorithms terminate if
the gap of the objectives between the last $10$ iterations is less
than $\epsilon\leq10^{-5}$.}
\label{tab:runtime}\centering %
\begin{tabular}{c|c||c||c|c|c|c|c|c}
\hline
\multicolumn{3}{c||}{$N_{\mathtt{T}}$ } & $2$  & $4$  & $6$  & $8$  & $10$  & $12$ \tabularnewline
\hline
\multirow{6}{*}{\parbox{1cm} {$N_{\mathtt{R}}=2$ $K_{\mathtt{D}}=2$\\ $K_{\mathtt{U}}=2$
}} & \multirow{3}{*}{i.i.d channel model} & Algorithm 1 (SDPT3)  & 2.61  & 3.74  & 5.61  & 9.46  & 15.14  & 17.92 \tabularnewline
\cline{3-9}
 &  & Algorithm 2 (SeDuMi) & 1.43 & 2.63  & 3.77  & 6.66  & 11.54  & 14.69 \tabularnewline
\cline{3-9}
 &  & Algorithm 2 (MOSEK) & 0.089  & 0.26  & 0.45 & 1.09  & 2.68  & 3.38 \tabularnewline
\cline{2-9}
 & \multirow{3}{*}{\parbox{2.8cm} {realistic channel model \\ (given in Sec. \ref{sub:Spectral-Efficiency-Performance})}} & Algorithm 1 (SDPT3)  & 4.17 & 6.28 & 9.29 & 15.04 & 23.76 & 28.51\tabularnewline
\cline{3-9}
 &  & Algorithm 2 (SeDuMi) & 2.36 & 4.13 & 6.11 & 10.50 & 17.64 & 22.93\tabularnewline
\cline{3-9}
 &  & Algorithm 2 (MOSEK) & 0.21 & 0.61 & 1.01 & 2.47 & 5.09 & 6.66\tabularnewline
\hline
\hline
\multicolumn{3}{c||}{$K_{\mathtt{D}}$} & $2$  & $4$  & $6$  & $8$  & $10$  & $12$ \tabularnewline
\hline
\multirow{6}{*}{\parbox{1cm} {$N_{\mathtt{T}}=4$ $N_{\mathtt{R}}=2$\\ $K_{\mathtt{U}}=2$
}} & \multirow{3}{*}{i.i.d channel model} & Algorithm 1 (SDPT3) & 3.74 & 9.64 & 13.01 & 16.27 & 18.76 & 25.32\tabularnewline
\cline{3-9}
 &  & Algorithm 2 (SeDuMi) & 2.63 & 6.25 & 8.12 & 9.98 & 12.77 & 15.96\tabularnewline
\cline{3-9}
 &  & Algorithm 2 (MOSEK) & 0.26 & 1.24 & 1.66 & 2.52 & 3.09 & 3.92\tabularnewline
\cline{2-9}
 & \multirow{3}{*}{\parbox{2.8cm} {realistic channel model \\ (given in Sec. \ref{sub:Spectral-Efficiency-Performance})} } & Algorithm 1 (SDPT3) & 6.28 & 17.33 & 22.84 & 27.55 & 31.58 & 42.57\tabularnewline
\cline{3-9}
 &  & Algorithm 2 (SeDuMi) & 4.13 & 10.59 & 14.19 & 17.05 & 22.54 & 27.91\tabularnewline
\cline{3-9}
 &  & Algorithm 2 (MOSEK) & 0.61 & 2.24 & 2.90 & 4.34 & 5.24 & 7.08\tabularnewline
\hline
\end{tabular}
\end{table*}

\subsection{\label{sub:Spectral-Efficiency-Performance}Spectral Efficiency Performance}

We now evaluate the performance of the full-duplex system for more
realistic models. Particularly, we compare the achievable spectral
efficiency of the proposed beamformer designs for the full-duplex
system introduced in Section \ref{sec:SM_PF} with that of a traditional
half-duplex scheme having the relevant hardware configurations. In
fact, as mentioned earlier, the application with the most potential
for full-duplex technology in cellular systems is in small cells.
To quantify the potential benefit of the full-duplex transmission
considered in this paper, we evaluate the performance of the proposed
algorithms under the 3GPP LTE specifications for small cell deployments.
The general simulation parameters are taken from \cite{LTE:TS36814,earthproject:D2.2}
and listed in Table \ref{table:sysparams}. Without loss of generality,
per-user power constraints of users in the uplink transmission are
assumed to be equal, i.e., $\overline{q}_{\mathtt{U}_{j}}=\overline{q}$.
In particular, we consider two different settings of the transmit
power constraints in both directions: (i) $(P_{\mathtt{BS}},\overline{q})=(26\;\textrm{dBm},23\;\textrm{dBm})$
following the LTE 3GPP pico cell standard for outdoor \cite{LTE:TS36814}
and (ii) $(P_{\mathtt{BS}},\overline{q})=(10\;\textrm{dBm},10\;\textrm{dBm})$
according to the work of \cite{Duarte:Fullduplex:Experiment}. The
number of antennas at the BS is set to $6$, of which $4$ are used
for transmitting and $2$ for receiving, i.e., $N_{\mathtt{T}}=4$
and $N_{\mathtt{R}}=2$, respectively. All users in both directions
are randomly dropped in a circle area of a radius $r=100$ m, centered
at the full-duplex capable BS in an outdoor small cell scenario.

\begin{table}
\caption{Simulation parameters}
\centering%
\begin{tabular}{l||r}
\hline
Carrier frequency  & $2$GHz\tabularnewline
System bandwidth  & $10$MHz\tabularnewline
Thermal noise  & $-174$ dBm/Hz\tabularnewline
Receiver noise figure (at downlink users) & $9$ dB\tabularnewline
Receiver noise figure (at BS) & $5$ dB\tabularnewline
\hline
Maximum transmit power at BS ($P_{\mathtt{BS}}$) & $10$ or $26$ dBm\tabularnewline
Maximum transmit power per user ($\bar{q}$) & $10$ or $23$ dBm\tabularnewline
\hline
\end{tabular}
\label{table:sysparams}
\end{table}

\begin{figure}
\centering{\subfigure[Location of users of the simulation setup considered in Figs. \ref{fig:Sumrate-vs-SI} and \ref{fig:CDF-chan} \label{fig:caseUE22}]{
\includegraphics[width=0.45\columnwidth]{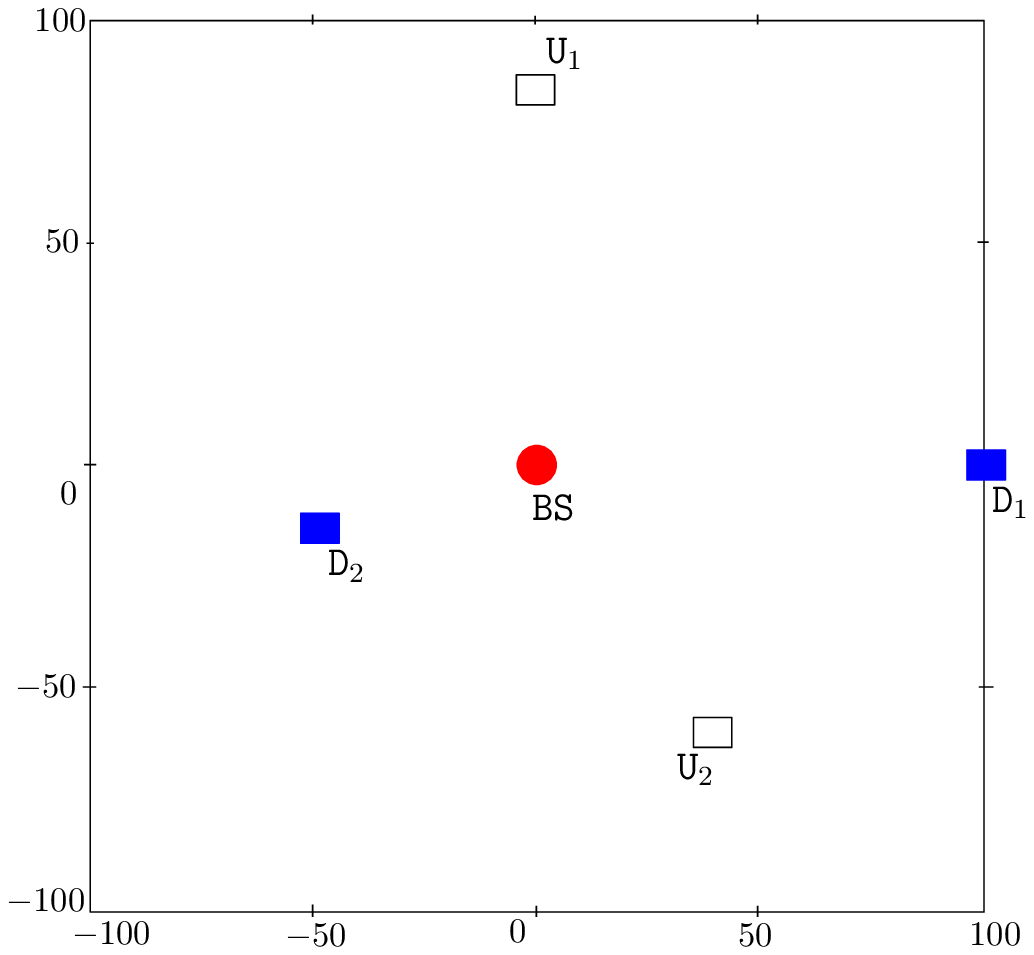}}\hspace{0.5cm}\centering\subfigure[Location of users of the simulation setup considered in Fig. \ref{fig:Sumrate-vs-Distance}\label{fig:caseUE11_location}]{\includegraphics[width=0.45\columnwidth]{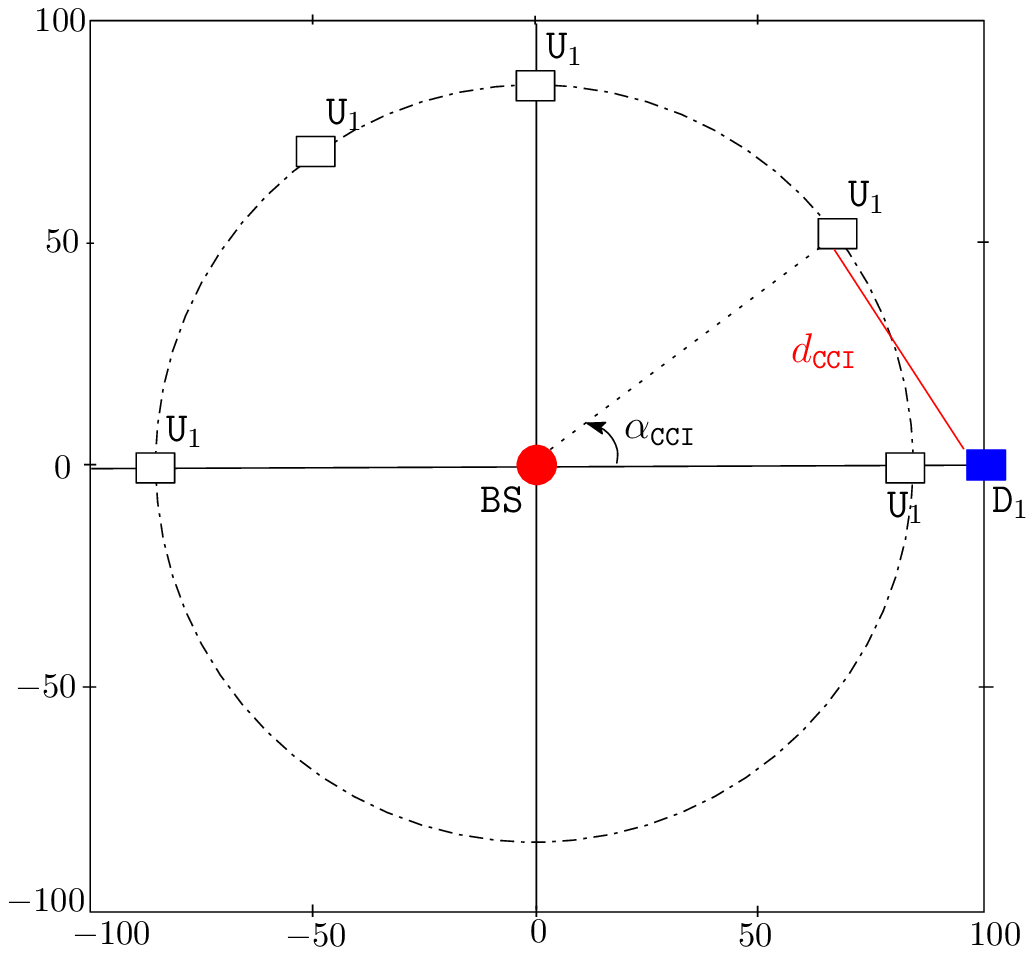}}}
\caption{\label{fig:setup}Location of users for the two specific simulation
settings considered in the numerical results section.}
\end{figure}

The channel vector from the BS to $\mathtt{D}_{i}$ is given by $\mathbf{h}_{\mathtt{D}_{i}}=\sqrt{\kappa_{\mathtt{D}_{i}}}\tilde{\mathbf{h}}_{\mathtt{D}_{i}}$
where $\tilde{\mathbf{h}}_{\mathtt{D}_{i}}$ follows $\mathcal{CN}(0,\mathbf{I})$
that denotes the small scale fading, and $\kappa_{\mathtt{D}_{i}}=10^{(-\textrm{PL}_{\mathtt{LOS}}/10)}$
represents the path loss, where $\textrm{PL}_{\mathtt{LOS}}$ is calculated
from a specific path loss model as shown in \eqref{eq:PL:LOS}. The
channel vector between the BS and $\mathtt{U}_{j}$ is generated in
the same way. For large scale fading, we adopt the path loss model
presented in \cite{earthproject:D2.2,LTE:TS36814}. More specifically,
downlink and uplink channels are assumed to experience the path loss
model for line of sight (LOS) communications as
\begin{equation}
\begin{array}{rl}
\textrm{PL}_{\mathtt{LOS}} & =103.8+20.9\log_{10}d\end{array}\label{eq:PL:LOS}
\end{equation}
where $\textrm{PL}_{\mathtt{LOS}}$ is in dB, $d$ is the distance
(in kilometers) between the BS and a specific user. Similarly, the
channel coefficient from $\mathtt{U}_{j}$ to $\mathtt{D}_{i}$ is
modeled as $g_{ji}=\sqrt{\kappa_{ji}}\tilde{g}_{ji}$ where $\tilde{g}_{ji}$
follows $\mathcal{CN}(0,1)$ and $\kappa_{ji}=10^{(-\textrm{PL}_{\mathtt{NLOS}}/10)}$
denotes the large scale fading. Since there is a high possibility
of obstructions between users deployed in an outdoor environment,
we assume that the channel from $\mathtt{U}_{j}$ to $\mathtt{D}_{i}$
encounters the path loss model for non-line-of-sight (NLOS) transmission.
That is, $\textrm{PL}_{\mathtt{NLOS}}$ (in dB) is written as
\begin{equation}
\textrm{PL}_{\mathtt{NLOS}}=145.4+37.5\log_{10}d_{\mathtt{CCI}}\label{eq:PL:NLOS}
\end{equation}
where $d_{\mathtt{CCI}}$ is now the distance (in kilometers) from
a user in the uplink transmission to another user in the downlink
direction. The self-interference channel model is mentioned in Subsection
\ref{sub:Convergence-and-Complexity}.

To have a fair comparison between the full-duplex and half-duplex
systems, we made the following assumptions. First, the BS of the half-duplex
counterpart is assumed to use all antennas in both downlink and uplink
transmissions, i.e., $N_{\mathtt{T}}+N_{\mathtt{R}}$. For the half-duplex
case, since the downlink and uplink transmissions are separated, and
thus the SEs of the downlink and uplink channels can be computed independently.
Specifically, we use the iterative water-filling algorithm introduced
in \cite{Yu:IterativeWF:MAC:2004} to find the optimal SE of the uplink
channel. Note that the problem of SE maximization in the downlink
direction is NP-hard which requires extremely high computational complexity
to find optimal solution \cite{Luo:SpectrumManagement:2008}. Herein,
we employ an efficient solution proposed in \cite{Nam:WSRM:SOCP:2012},
which was shown to be close optimal, to calculate the SE of the downlink
transmission. Then, the resulting SEs of the downlink and uplink channels
in the half-duplex counterpart are divided by $2$ since each of them
is assumed to share $50\%$ of the temporal resource \cite{Duarte:Fullduplex:Experiment}.
For the full-duplex case, the SEs of the downlink and uplink channels
are simply calculated by \eqref{eq:DL:SumRate:general} and \eqref{eq:UL:SumRate},
respectively, after achieving the solutions of the problem in \eqref{eq:Original:SE}.

\begin{figure}
\centering\subfigcapskip = 0cm \subfigure[Average spectral efficiency gain of downlink channel. \label{fig:DLSRvsSI}]{\includegraphics[width=1\columnwidth]{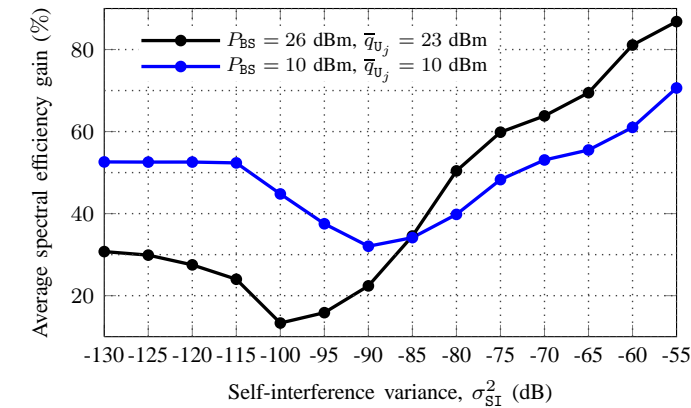}}
\vspace{0.35cm}\hspace{-0.4cm}\subfigcapskip = 0cm\subfigure[Average spectral efficiency gain of uplink channel. \label{fig:ULSRvsSI}]{\includegraphics[width=1\columnwidth]{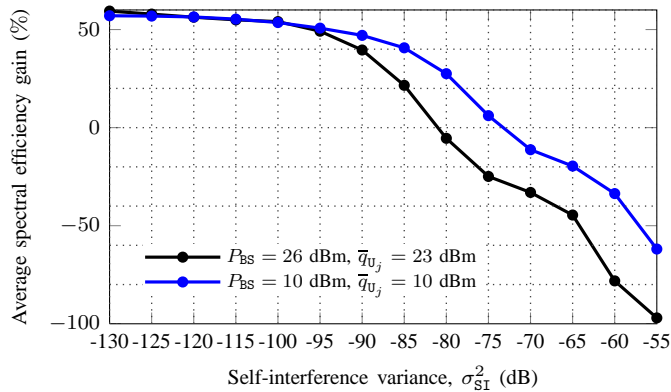}}
 \vspace{0.35cm}\hspace{0.1cm}\subfigcapskip = 0cm\subfigure[Average spectral efficiency gain of entire system. \label{fig:TotalSRvsSI}]{\includegraphics[width=1\columnwidth]{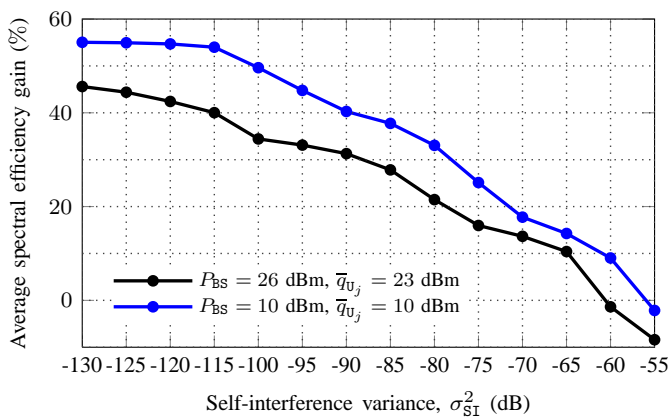}}
\caption{\label{fig:Sumrate-vs-SI}Average spectral efficiency gain ($\%$)
versus $\sigma_{\mathtt{SI}}^{2}$ (dB) for the simulation scenario
shown in Fig. \ref{fig:caseUE22}. }
\end{figure}

Fig. \ref{fig:Sumrate-vs-SI} depicts the SE gains in percentage of
the full-duplex system over the half-duplex one as a function of $\sigma_{\mathtt{SI}}^{2}$
for the scenario as shown in Fig. \ref{fig:caseUE22}. A general observation
is that full-duplex transmission can significantly improve the spectral
efficiency of the half-duplex one when the self-interference is substantially
suppressed. Specifically, as shown in Fig. \ref{fig:TotalSRvsSI},
the total SE gain of the full-duplex system is $45.6\%$ and $55\%$
for the cases $(P_{\mathtt{BS}},\overline{q})=(26\;\textrm{dBm},23\;\textrm{dBm})$
and $(P_{\mathtt{BS}},\overline{q})=(10\;\textrm{dBm},10\;\textrm{dBm})$
at $\sigma_{\mathtt{SI}}^{2}=-130$ dB, respectively. However, when
$\sigma_{\mathtt{SI}}^{2}=-55$ dB, the half-duplex system performs
better than the full-duplex one for both cases of transmit power constraint.
This observation simply means that the self-interference cancellation
mechanism should be efficient enough for the full-duplex system to
compete against the half-duplex counterpart. In addition,\textit{
the simulation results also indicate that the self-interference needs
to be canceled at least $75$ dB (i.e., $\sigma_{\mathtt{SI}}^{2}<-75$
dB) for the case $(P_{\mathtt{BS}},\overline{q})=(10\;\textrm{dBm},10\;\textrm{dBm})$
and at least $83$ dB (i.e., $\sigma_{\mathtt{SI}}^{2}<-83$ dB) for
the case $(P_{\mathtt{BS}},\overline{q})=(26\;\textrm{dBm},23\;\textrm{dBm})$
for the full-duplex system to attain better SE in both downlink and
uplink transmissions, compared to the half-duplex one. These requirements
can be achieved by a recent advanced SI cancellation technique reported
in \cite{Dinesh:FullduplexRadio:Sigcom:2013}.}

To obtain more insights into the performance of the full-duplex system,
we also study the gains of the downlink and uplink channels separately
in Figs. \ref{fig:DLSRvsSI} and \ref{fig:ULSRvsSI}, respectively.
We can see that, while the SE of the uplink transmission of the full-duplex
system is always deteriorated as $\sigma_{\mathtt{SI}}^{2}$ increases,
that of the downlink channel decreases until a certain value of $\sigma_{\mathtt{SI}}^{2}$
($-100$ dB and $-90$dB for $(P_{\mathtt{BS}},\overline{q})=(26\;\textrm{dBm},23\;\textrm{dBm})$
and $(P_{\mathtt{BS}},\overline{q})=(10\;\textrm{dBm},10\;\textrm{dBm})$,
respectively) and increases after that. The degradation on the SE
of the uplink channel is obvious and due to the fact that a large
value of $\sigma_{\mathtt{SI}}^{2}$ results in a greater amount of
self-interference power being added to the background noise. To explain
different trends in the SE of the downlink channel, we first recall
that the main goal of the proposed designs is to maximize the total
SE of the full-duplex system, i.e., jointly optimizing both uplink
and downlink transmissions. When the SI is quite small, the joint
optimization schemes slightly reduce the actual transmit power of
the downlink channel to maintain the SE of the uplink channel. For
a large value of $\sigma_{\mathtt{SI}}^{2}$, the self-interference
is comparable or even dominates the desired signals of the users in
the uplink channel. Hence, data detection for uplink users becomes
more erroneous, incredibly deteriorating the uplink performance. For
such a case, the total SE of the full-duplex system is mostly determined
by the downlink transmission since the SE of the uplink channel is
extremely low. Thus, it is better to reduce the transmit power in
the uplink channel and concentrate on maximizing the SE of the downlink
channel. As a result, the SE of the uplink channel greatly declined.
Specifically, the SE of the uplink direction of the full-duplex system
is remarkably smaller than that of the half-duplex one as $\sigma_{\mathtt{SI}}^{2}\geq-80$
and $\sigma_{\mathtt{SI}}^{2}\geq-70$ for $(P_{\mathtt{BS}},\overline{q})=(26\;\textrm{dBm},23\;\textrm{dBm})$
and $(P_{\mathtt{BS}},\overline{q})=(10\;\textrm{dBm},10\;\textrm{dBm})$,
respectively. It is worth noting that a reduction in the transmit
power of users in uplink channel results in a decrease in the CCI.
This explains the increment of the SE gain in the downlink transmission
as $\sigma_{\mathtt{SI}}^{2}$ is greater than a certain threshold.
An interesting observation from Fig. \ref{fig:TotalSRvsSI} is that
the SE gain of the full-duplex system is higher when the maximum transmit
power is smaller. This is due to the fact that smaller maximum transmit
powers create a smaller amount of self-interference as well as CCI.

\begin{figure}
\centering { \includegraphics[width=1\columnwidth]{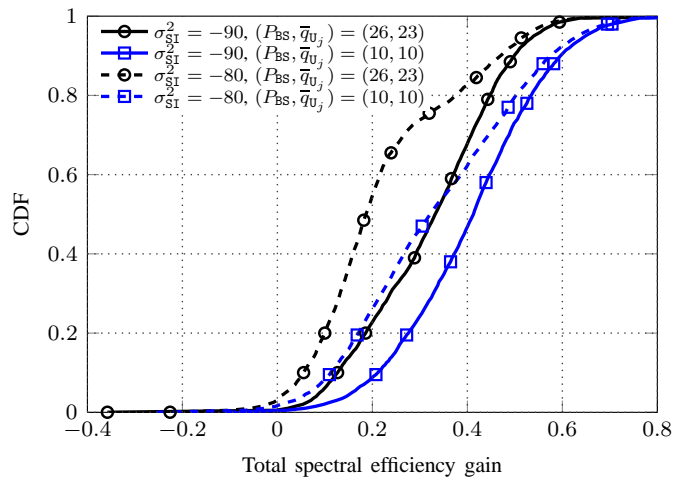}}\vspace{0pt}
\caption{\label{fig:CDF-chan} CDF of total spectral efficiency gains for $5000$
random channel realizations for the scenario shown in Fig. \ref{fig:caseUE22}.
The unit of $\sigma_{\mathtt{SI}}^{2}$ is dB and that of $P_{\mathtt{BS}}$
and $\overline{q}$ is dBm. }
\end{figure}

In Fig. \ref{fig:CDF-chan}, we show cumulative distribution function
(CDF) of the total SE gain of the full-duplex for the scenario in
Fig. \ref{fig:caseUE22}. Obviously, for the same power setting, a
smaller value of $\sigma_{\mathtt{SI}}^{2}$ results in better SE
gain. On the other hand, for the same $\sigma_{\mathtt{SI}}^{2}$,
a lower transmit power yields better SE gain. These observations are
consistent with the observation in Fig. \ref{fig:TotalSRvsSI}.

\begin{figure}
\centering\subfigcapskip = 0cm \subfigure[CDF of average spectral efficiency gain of downlink channel. \label{fig:CDF_DL}]{\includegraphics[width=1\columnwidth]{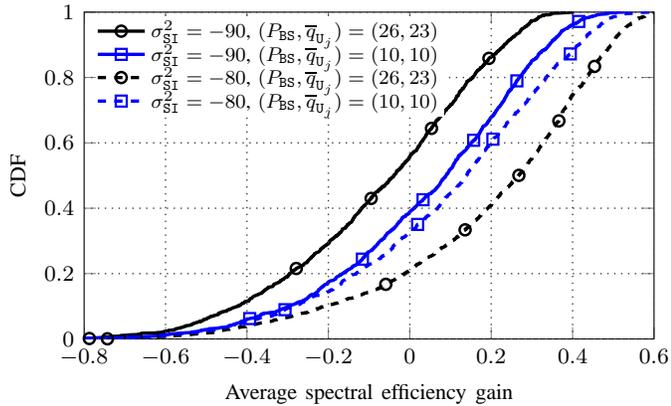}}
	\vspace{0.35cm}\hspace{0.1cm}\subfigcapskip = 0cm\subfigure[CDF of average spectral efficiency gain of uplink channel. \label{fig:CDF_UL}]{\includegraphics[width=1\columnwidth]{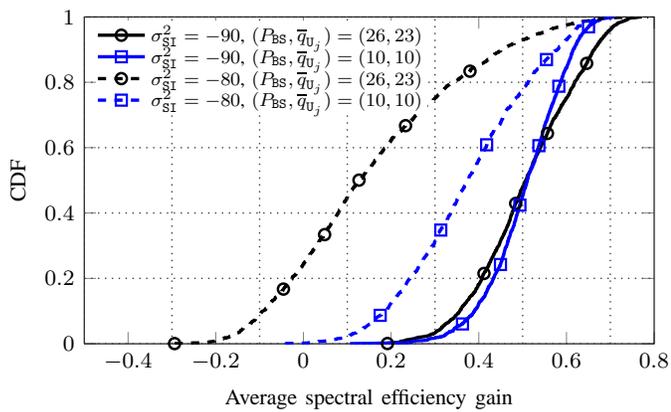}}
 \vspace{0.35cm}\hspace{0.1cm}\subfigcapskip = 0cm\subfigure[CDF of average spectral efficiency gain of entire system. \label{fig:CDF_Total}]{\includegraphics[width=1\columnwidth]{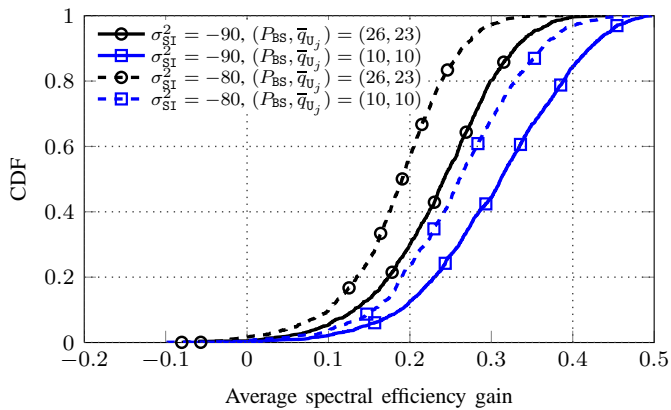}}
\caption{\label{fig:CDF} CDF of average spectral efficiency gains for $1000$
random topologies. The simulation scenario parameters are $K_{\mathtt{D}}=K_{\mathtt{U}}=2$,
$N_{\mathtt{T}}=4$ and $N_{\mathtt{R}}=2$. The users are uniformly
dropped in a circle area of a radius $r=100$ meters centered at the
BS at random. The unit of $\sigma_{\mathtt{SI}}^{2}$ is dB and that
of $P_{\mathtt{BS}}$ and $\overline{q}$ is dBm.}
\end{figure}

The performance of the full-duplex is further explored in the next
numerical experiment, in which we study the CDF of the average SE
gain of the full-duplex system for a number of random topologies.
The results in Fig. \ref{fig:CDF} are plotted for $1000$ topologies,
where all users are uniformly distributed in a circle area of a radius
$r=100$ meters centered at the BS. For each topology, the spectral
efficiency gain is averaged over $500$ random channel realizations.
As can be seen in Fig. \ref{fig:CDF_Total}, the total average SE
of full-duplex systems are higher than that of the half-duplex one
for most of the topologies. For example, the SE gains are larger than
$20\%$ and $28\%$ for the power settings $(P_{\mathtt{BS}},\overline{q})=(26\;\textrm{dBm},23\;\textrm{dBm})$
and $(P_{\mathtt{BS}},\overline{q})=(10\;\textrm{dBm},10\;\textrm{dBm})$,
respectively for a half of the simulated topologies at $\sigma_{\mathtt{SI}}^{2}=-80$
dB. Not surprisingly, the SE gain of the downlink channel is rather
sensitive to topologies which determine the degree of CCI. On the
other hand, positions of users have a small impact on the SE of the
uplink transmission when $\sigma_{\mathtt{SI}}^{2}=-90$ dB. The reason
is that the self-interference in this case is relatively lower than
the received signal strength for most of the topologies. However,
the situation dramatically changes as $\sigma_{\mathtt{SI}}^{2}$
increases to $-80$ dB, where more dependency between topology and
SE gain is observed. Thus, the number of scenarios that can yield
a received signal strength higher than the SI power is reduced for
a larger value of $\sigma_{\mathtt{SI}}^{2}$.

\begin{figure}
\centering \subfigcapskip = 0cm\subfigure[Average spectral efficiency of downlink channel.  \label{fig:DLvsdist}]{ \includegraphics[width=1\columnwidth]{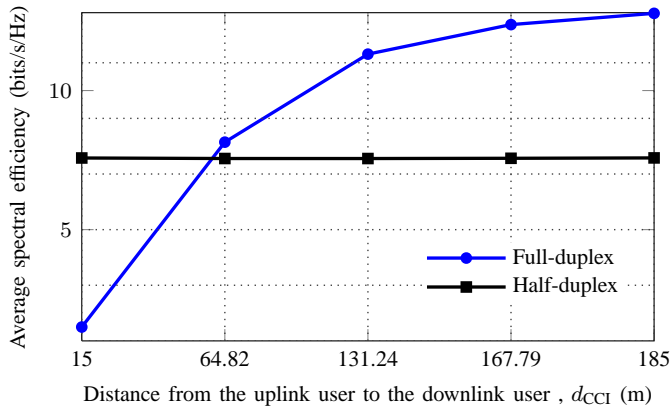}}
\vspace{0.35cm}\hspace{0.1cm}\subfigcapskip = 0cm \subfigure[Average spectral efficiency of uplink channel. \label{fig:ULvsdist}]{\includegraphics[width=1\columnwidth]{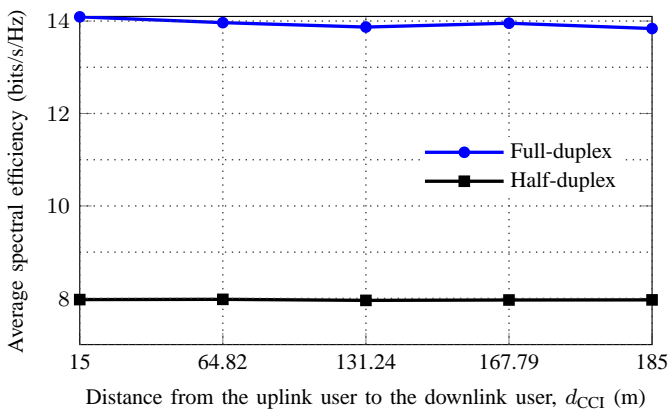}}
\vspace{0.35cm}\hspace{0.1cm}\subfigcapskip = 0cm \subfigure[Average spectral efficiency of entire system. \label{fig:Totalvsdist}]{\includegraphics[width=1\columnwidth]{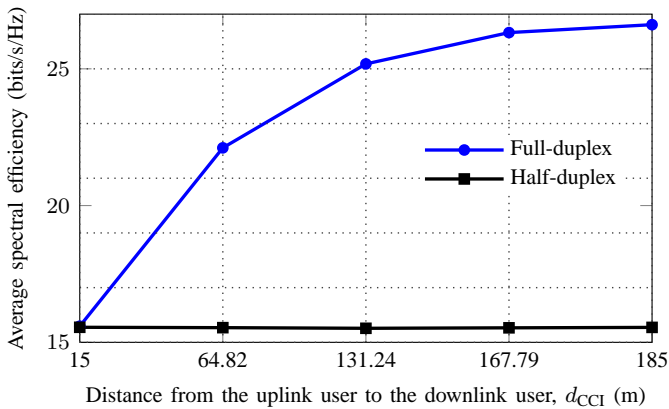}}
\caption{\label{fig:Sumrate-vs-Distance}Average spectral efficiency versus
distance from the uplink user to downlink one, $d_{\mathtt{CCI}}$.
In this setup, $\sigma_{\mathtt{SI}}^{2}=-100$ dB, $P_{\mathtt{BS}}=26$
dBm and $\overline{q}_{\mathtt{U}_{j}}=23$ dBm. The distance of the
BS and users, i.e., $\mathtt{D}_{1}$, and $\mathtt{U}_{1}$ are set
at $r$ and $0.85r$, respectively. The position of $\mathtt{D}_{1}$
is fixed while $\mathtt{U}_{1}$ moves on a circle with radius $0.85r$
as shown in Fig. \ref{fig:caseUE11_location}.}
\end{figure}

\begin{figure}
\centering { \includegraphics[width=1\columnwidth]{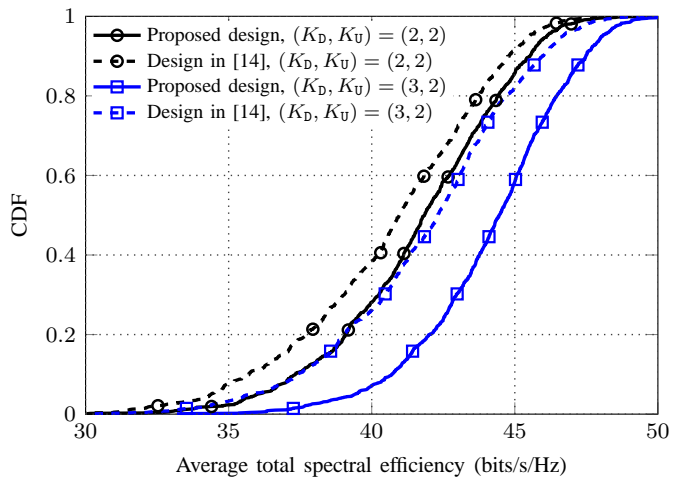}}\caption{\label{fig:CDF-CCInoCCI}CDF of average total spectral efficiency
of the proposed design and the design with no consideration of CCI
in \cite{Dan:Fullduplex:MIMO:2013} for $1000$ random topologies.
In the simulation setting, $\sigma_{\mathtt{SI}}^{2}=-100$ dB, $P_{\mathtt{BS}}=26$
dBm and $\overline{q}_{\mathtt{U}_{j}}=23$ dBm.}
\end{figure}

Next, we study the impact of co-channel interference on the SE of
the full-duplex system. For this purpose, we fix $\sigma_{\mathtt{SI}}^{2}$
at $-100$ dB, and consider a setting shown in Fig. \ref{fig:caseUE11_location}.
In this simulation setup, we vary the distance between $\mathtt{U}_{1}$
and $\mathtt{D}_{1}$, denoted by $d_{\mathtt{CCI}}$, and plot the
resulting SEs of the full-duplex system in Fig. \ref{fig:Sumrate-vs-Distance}.
Each value of $d_{\mathtt{CCI}}$ on the x-axis of Fig. \ref{fig:Sumrate-vs-Distance}
corresponds to a position of $\mathtt{U}_{1}$, while $\mathtt{D}_{1}$
is held fixed. We observe that the spectral efficiency of $\mathtt{D}_{1}$
increases as $\mathtt{U}_{1}$ moves far away from $\mathtt{D}_{1}$.
Especially when the two users are close (e.g., $d_{\mathtt{CCI}}<64.82$
m), the performance of the full-duplex downlink transmission can be
worse than that of the half-duplex one. The reason is straightforward
since decreasing $d_{\mathtt{CCI}}$ leads to an increase in CCI which
then degrades the SE of the downlink channel. On the other hand, the
location of $\mathtt{U}_{1}$ has a small impact on the SE of the
uplink transmission for a fixed small value of $\sigma_{\mathtt{SI}}^{2}$.
The results in Fig. \ref{fig:Sumrate-vs-Distance} indicate that the
CCI is a critical factor that needs to be controlled for successful
deployment of full-duplex systems.

In the final numerical experiment we plot the CDF of the average total
SE of the full-duplex system with and without accounting for the CCI.
The problem of beamformer design without taking CCI into account was
studied in \cite{Dan:Fullduplex:MIMO:2013}. The curves in Fig. \ref{fig:CDF-CCInoCCI}
are obtained from 1000 random topologies. For each topology, the average
total SE is calculated over $500$ random channel realizations. It
is obvious that the proposed designs in this paper outperform the
one with no CCI in \cite{Dan:Fullduplex:MIMO:2013} as expected. For
instance, the proposed designs attain 2 bits/s/Hz of total SE higher
than the scheme in \cite{Dan:Fullduplex:MIMO:2013} for approximately
$60\%$ of the simulated topologies when $K_{\mathtt{D}}=3$ and $K_{\mathtt{U}}=2$.
As the total number of users is reduced, the SE becomes smaller due
to a decrease in the available multiuser diversity gain. \vspace{-10pt}

\section{\label{sec:Conclusions}Conclusion and Future Work }

In this paper we have devised a beamforming scheme for a full-duplex
system, in which a full-duplex capable BS communicates with multiple
half-duplex users in the downlink and uplink channels simultaneously.
In particular, we have considered the problem of joint SE maximization
of downlink and uplink transmissions under some power constraints.
First, the design problem is formulated as a rank constrained optimization
one, and then the rank relaxation technique is applied. However, the
relaxed problem is still nonconvex. To solve this problem we have
proposed two iterative algorithms, one based on the concept of the
FW algorithm and the other based on the framework of SPCA method.
The idea of both proposed methods is to approximate the nonconvex
problem by a convex formulation in each iteration. While the first
approach needs to solve a sequence of MAXDET programs, the second
one relies on solving a series of SDPs. We have carried out several
numerical experiments under 3GPP LTE small cell setups to evaluate
the SE performance of the full-duplex scheme. It has been shown that
the SE of the full-duplex system is remarkably larger than that of
the half-duplex one as the capability of current SI cancellation schemes
is efficient. Our work has proved that the full-duplex transmission
is a promising technique to improve the SE of small cell wireless
communications systems.

The work considered in this paper also opens several possibilities
for future research. First, more efficient designs of self-interference
cancellation for full-duplex MIMO systems are of critical importance.
In addition to distributed algorithms for multiple small cell setups
as mentioned earlier, a mechanism which can accurately measure the
CCI at users in the downlink channel is required. When many users
are active in the downlink and uplink channels, a CCI-aware user scheduling
scheme which can control the CCI is a good solution to the full-duplex
systems. This allows us to exploit the multiuser-diversity gain in
both directions. Furthermore, since the uplink performance of the
full-duplex system is significantly reduced, even worse than the half-duplex
one due to a large amount of self-interference, a mechanism to control
the fairness among users needs to be proposed. For example, we can
additionally impose a rate constraint on the SE of the uplink channel.
The future research can also include an efficient algorithm to switch
between full-duplex and half-duplex systems. Since the downlink and
uplink channels operate at the same time, some traditional MAC protocols,
which are dedicated to current half-duplex systems, need to be redesigned.
These interesting problems call for more comprehensive studies, and
thus are beyond the scope of this paper.

\appendix[Proof of Convergence]

In this appendix we adopt the techniques from \cite{Beck:SCA:2010}
to prove the convergence of Algorithms \ref{algo:iterativeMAXDET}
and \ref{algo:IterativeSDP} (i.e., the iterative MAXDET-based algorithm
and the iterative SDP-based algorithm, respectively) to a KKT point.
Let us start with the convergence proof of Algorithm \ref{algo:iterativeMAXDET}.
First, we note that the affine majorization in \eqref{eq:affine}
has the following two important properties which are the key to show
the convergence to a KKT point of Algorithm \ref{algo:iterativeMAXDET}
\begin{eqnarray}
g^{(n)}(\mathbf{Q}^{(n)},\mathbf{q}^{(n)}) & = & g(\mathbf{Q}^{(n)},\mathbf{q}^{(n)}),\label{eq:tightness}\\
\nabla g^{(n)}(\mathbf{Q}^{(n)},\mathbf{q}^{(n)}) & = & \nabla g(\mathbf{Q}^{(n)},\mathbf{q}^{(n)})\label{eq:grad}
\end{eqnarray}
where property \eqref{eq:tightness} means that the inequality in
\eqref{eq:affine} is tight when $(\mathbf{Q},\mathbf{q})=(\mathbf{Q}^{(n)},\mathbf{q}^{(n)})$
and property \eqref{eq:grad} is obvious due to the first order approximation.
Note that the gradient in \eqref{eq:grad} is with respect to $\mathbf{Q}$
and $q$. To proceed further, let $\mathcal{S}$ denote the feasible
set of \eqref{eq:MAXDET}, i.e., the set of $\mathbf{Q}$ and $\mathbf{q}$
that satisfy the constraint \eqref{eq:SE:c1}, \eqref{eq:SE:c2} and
\eqref{eq:SE:c3}. We note that $\mathcal{S}$ is a compact convex
set. Further let $u^{(n+1)}$ be is the obtained optimal objective
of \eqref{eq:MAXDET} at iteration $n+1$. According to the updating
rule in Algorithm \ref{algo:iterativeMAXDET}, we can derive the following
inequalities \begin{IEEEeqnarray}{rCl}\IEEEyesnumber\label{eq:proof:algorithm1}
u^{(n+1)} & = & h(\mathbf{Q}^{(n+1)},\mathbf{q}^{(n+1)})-g^{(n)}(\mathbf{Q}^{(n+1)},\mathbf{q}^{(n+1)})\\  & = & \underset{(\mathbf{Q},\mathbf{q})\in\mathcal{S}}{\max}\ h(\mathbf{Q},\mathbf{q})-g^{(n)}(\mathbf{Q},\mathbf{q})\\  & \geq & h(\mathbf{Q}^{(n)},\mathbf{q}^{(n)})-g^{(n)}(\mathbf{Q}^{(n)},\mathbf{q}^{(n)})\label{eq:proofa}\\  & = & h(\mathbf{Q}^{(n)},\mathbf{q}^{(n)})-g(\mathbf{Q}^{(n)},\mathbf{q}^{(n)})\label{eq:proofb}\\  & \geq & h(\mathbf{Q}^{(n)},\mathbf{q}^{(n)})-g^{(n-1)}(\mathbf{Q}^{(n)},\mathbf{q}^{(n)})=u^{(n)}\label{eq:proofc}\IEEEeqnarraynumspace \end{IEEEeqnarray}where \eqref{eq:proofa} follows from the fact that the objective
at the optimal solution is greater than the one at any feasible solution,
i.e., $f(\mathbf{x}^{\star})=\underset{\mathbf{x}\in\mathcal{X}}{\max}f(\mathbf{x})\geq f(\mathbf{x}_{0})$
where $\mathbf{x}^{\star}$ and $\mathbf{x}_{0}$ are an optimal solution
and any feasible solution, respectively, \eqref{eq:proofb} is due
to \eqref{eq:tightness}, \eqref{eq:proofc} is due to the affine
majorization in \eqref{eq:affine}. In fact, we have shown that the
sequence $\{u^{(n)}\}$ in \emph{nondecreasing}. Furthermore, the
value of $\{u^{(n)}\}$ is bounded above due to the limited transmit
power, and thus it is guaranteed to converge. We note that the function
$f(\mathbf{X})=\log\det(\mathbf{X})$ is differentiable and strictly
concave on $\mathbf{X}\succ\mathbf{0}$ \cite[Section 3.1]{Boyd:ConvexOpt:2004}.
Since $\mathcal{S}$ is a compact convex set, the objective is then
shown to be strongly concave on $\mathcal{S}$ due to \cite[Lemma 3.1]{Beck:SCA:2010}.
As a result, the sequence $(\mathbf{Q}^{(n)},\mathbf{q}^{(n)})$ converges
to an accumulation point denoted by $(\mathbf{Q}^{\ast},\mathbf{q}^{\ast})$.
To establish the convergence to a KKT point, we first introduce the
set of dual variables for the constraints in \eqref{eq:MAXDET} which
is listed in Table \ref{tab:dualvariable}.

\begin{table}[b]
\caption{Constraints and their corresponding dual variables}
\label{tab:dualvariable}
\centering
\begin{tabular}{c|c}
\hline
Constraints & Dual variables\tabularnewline
\hline
$0\leq q_{\mathtt{U}_{j}}$ & $\lambda_{\mathtt{U}_{j}}$\tabularnewline
\hline
$q_{\mathtt{U}_{j}}\leq\overline{q}_{\mathtt{U}_{j}}$ & $\tilde{\lambda}_{\mathtt{U}_{j}}$\tabularnewline
\hline
$\sum_{i=1}^{K_{\mathtt{D}}}\tr(\mathbf{Q}_{\mathtt{D}_{i}})\leq P_{\mathtt{BS}}$ & $\mu$\tabularnewline
\hline
$\mathbf{Q}_{\mathtt{D}_{i}}\succeq0$ & $\mathbf{Z}_{\mathtt{D}_{i}}$\tabularnewline
\hline
\end{tabular}
\end{table}

It is easy to check that the Slater\textquoteright{}s condition holds
for the convex program at all iterations of Algorithm 1. Thus, the
KKT conditions are necessary and sufficient for optimality \cite[Section 5.5]{Boyd:ConvexOpt:2004}.
With the dual variables introduced in Table \ref{tab:dualvariable},
the KKT conditions of the optimal value at iteration $n$ (see \cite{Boyd:ConvexOpt:2004}
for more details) are given as\begin{multline}
\nabla_{\mathbf{Q}_{\mathtt{D}_{i}}}h(\mathbf{Q}^{(n)},\mathbf{q}^{(n)})-\nabla_{\mathbf{Q}_{\mathtt{D}_{i}}}g^{(n)}(\mathbf{Q}^{(n)},\mathbf{q}^{(n)})\\-\mu\mathbf{I}+\mathbf{Z}_{\mathtt{D}_{i}}\;=\;\mathbf{0},\ \forall i=1,\ldots,K_{\mathtt{D}},\label{eq:stationary1}
\end{multline}
\begin{multline}
\partial_{q_{\mathtt{U}_{j}}}h(\mathbf{Q}^{(n)},\mathbf{q}^{(n)})-\partial_{q_{\mathtt{U}_{j}}}g^{(n)}(\mathbf{Q}^{(n)},\mathbf{q}^{(n)})\\+\lambda_{\mathtt{U}_{j}}-\tilde{\lambda}_{\mathtt{U}_{j}}\;=\;0,\ \forall j=1,\ldots,K_{\mathtt{U}},\label{eq:stationary2}
\end{multline}
\begin{equation}
\lambda_{\mathtt{U}_{j}}q_{\mathtt{U}_{j}}^{(n)}=0;\ \tilde{\lambda}_{\mathtt{U}_{j}}(q_{\mathtt{U}_{j}}^{(n)}-\overline{q}_{\mathtt{U}_{j}})=0,\ \forall j=1,\ldots,K_{\mathtt{U}},\label{eq:cmplslack1}
\end{equation}
\begin{equation}
\tr(\mathbf{Q}_{\mathtt{D}_{i}}^{(n)}\mathbf{Z}_{\mathtt{D}_{i}})=0,\ \forall i=1,\ldots,K_{\mathtt{D}},\label{eq:cmplslack2}
\end{equation}
\begin{equation}
\mu\bigl(\sum_{i=1}^{K_{\mathtt{D}}}\tr(\mathbf{Q}_{\mathtt{D}_{i}}^{(n)})-P_{\mathtt{BS}}\bigr)=0.\label{eq:cmplslack3}
\end{equation}
Due to property \eqref{eq:grad}, we can replace $\nabla_{\mathbf{Q}_{\mathtt{D}_{i}}}g^{(n)}(\mathbf{Q}^{(n)},\mathbf{q}^{(n)})$
and $\partial_{q_{\mathtt{U}_{j}}}g^{(n)}(\mathbf{Q}^{(n)},\mathbf{q}^{(n)})$
by $\nabla_{\mathbf{Q}_{\mathtt{D}_{i}}}g(\mathbf{Q}^{(n)},\mathbf{q}^{(n)})$
and $\partial_{q_{\mathtt{U}_{j}}}g(\mathbf{Q}^{(n)},\mathbf{q}^{(n)})$
on convergence (i.e., as $n\to\infty$), respectively. Thus,\begin{multline}
\nabla_{\mathbf{Q}_{\mathtt{D}_{i}}}h(\mathbf{Q}^{(n)},\mathbf{q}^{(n)})-\nabla_{\mathbf{Q}_{\mathtt{D}_{i}}}g(\mathbf{Q}^{(n)},\mathbf{q}^{(n)})\\-\mu\mathbf{I}+\mathbf{Z}_{\mathtt{D}_{i}}\;=\;\mathbf{0},\ \forall i=1,\ldots,K_{\mathtt{D}},\label{eq:stationary1:origin}
\end{multline}
\begin{multline}
\partial_{q_{\mathtt{U}_{j}}}h(\mathbf{Q}^{(n)},\mathbf{q}^{(n)})-\partial_{q_{\mathtt{U}_{j}}}g(\mathbf{Q}^{(n)},\mathbf{q}^{(n)})\\+\lambda_{\mathtt{U}_{j}}-\tilde{\lambda}_{\mathtt{U}_{j}}\;=\;0,\ \forall j=1,\ldots,K_{\mathtt{U}}.\label{eq:stationary2:origin}
\end{multline}
It is straightforward to see that the set of equations in \eqref{eq:cmplslack1}-\eqref{eq:stationary2:origin}
are actually the KKT conditions for the problem \eqref{eq:DCform}
and thus completes the proof. We note that the KKT conditions for
the convex program after convergence are also the necessary ones for
local optimality of the problem \eqref{eq:DCform}. Indeed since $(\mathbf{Q}^{\ast},\mathbf{q}^{\ast})$
is an optimal solution to the convex program at convergence, it satisfies
\cite[Section 2.1]{Bertsekas:NonlinearProgramming:2003}
\begin{multline}
\bigl\langle\nabla g^{(\infty)}(\mathbf{Q}^{\ast},\mathbf{q}^{\ast})-\nabla h(\mathbf{Q}^{\ast},\mathbf{q}^{\ast}),(\mathbf{Q}^{\prime},\mathbf{q}^{\prime})\\-(\mathbf{Q}^{\ast},\mathbf{q}^{\ast})\bigr\rangle\geq0\ \text{for all}\ (\mathbf{Q}^{\prime},\mathbf{q}^{\prime})\in\mathcal{S}\label{eq:1stcond:app}
\end{multline} where $\left\langle ,\right\rangle $ stands for the inner product
of the arguments, i.e., $\bigl\langle\mathbf{X},\mathbf{Y}\bigr\rangle=\tr(\mathbf{X}^{H}\mathbf{Y})$,
the subtraction in \eqref{eq:1stcond:app} is element-wise, and the
gradient is with respect to $\mathbf{Q}$ and $\mathbf{q}$. As mentioned
previously, we can replace $\nabla g^{(\infty)}(\mathbf{Q}^{\ast},\mathbf{q}^{\ast})$
by $\nabla g(\mathbf{Q}^{\ast},\mathbf{q}^{\ast})$, and thus \eqref{eq:1stcond:app}
becomes\begin{multline}
\bigl\langle\nabla g(\mathbf{Q}^{\ast},\mathbf{q}^{\ast})-\nabla h(\mathbf{Q}^{\ast},\mathbf{q}^{\ast}),(\mathbf{Q}^{\prime},\mathbf{q}^{\prime})\\-(\mathbf{Q}^{\ast},\mathbf{q}^{\ast})\bigr\rangle\geq0\ \text{for all}\ (\mathbf{Q}^{\prime},\mathbf{q}^{\prime})\in\mathcal{S}\label{eq:1stcond:org}
\end{multline}which is the first order necessary conditions for local optimality
of the problem \eqref{eq:DCform} \cite[Section 2.1]{Bertsekas:NonlinearProgramming:2003}.

The proof of Algorithm \ref{algo:IterativeSDP} follows the same spirit.
As mentioned earlier for the convex approximation in \eqref{eq:c1:CVX:UB},
$F(t_{\mathtt{D}_{i}},\beta_{\mathtt{D}_{i}},\psi_{\mathtt{D}_{i}}^{(n)})=f(t_{\mathtt{D}_{i}},\beta_{\mathtt{D}_{i}})$
when $\psi_{\mathtt{D}_{i}}^{(n)}=t_{\mathtt{D}_{i}}/\beta_{\mathtt{D}_{i}}$,
that is
\begin{equation}
F(t_{\mathtt{D}_{i}},\beta_{\mathtt{D}_{i}},\psi_{\mathtt{D}_{i}}^{(n)})|_{\psi_{\mathtt{D}_{i}}^{(n)}=t_{\mathtt{D}_{i}}/\beta_{\mathtt{D}_{i}}}=t_{\mathtt{D}_{i}}\beta_{\mathtt{D}_{i}}=f(t_{\mathtt{D}_{i}},\beta_{\mathtt{D}_{i}}).
\end{equation}
Furthermore, we also have
\begin{equation}
\begin{array}{rl}
\left.\frac{\partial F(t_{\mathtt{D}_{i}},\beta_{\mathtt{D}_{i}},\psi_{\mathtt{D}_{i}}^{(n)})}{\partial t_{\mathtt{D}_{i}}}\right|_{\psi_{\mathtt{D}_{i}}^{(n)}=t_{\mathtt{D}_{i}}/\beta_{\mathtt{D}_{i}}} & =\left.{\displaystyle \frac{1}{\psi_{\mathtt{D}_{i}}^{(n)}}}t_{\mathtt{D}_{i}}\right|_{\psi_{\mathtt{D}_{i}}^{(n)}=t_{\mathtt{D}_{i}}/\beta_{\mathtt{D}_{i}}}\\
\\
 & =\beta_{\mathtt{D}_{i}}={\displaystyle \frac{\partial f(t_{\mathtt{D}_{i}},\beta_{\mathtt{D}_{i}})}{\partial t_{\mathtt{D}_{i}}}}
\end{array}
\end{equation}
 and
\begin{equation}
\left.\frac{\partial F(t_{\mathtt{D}_{i}},\beta_{\mathtt{D}_{i}},\psi_{\mathtt{D}_{i}}^{(n)})}{\partial\beta_{\mathtt{D}_{i}}}\right|_{\psi_{\mathtt{D}_{i}}^{(n)}=t_{\mathtt{D}_{i}}/\beta_{\mathtt{D}_{i}}}=\frac{\partial f(t_{\mathtt{D}_{i}},\beta_{\mathtt{D}_{i}})}{\partial\beta_{\mathtt{D}_{i}}}.
\end{equation}
Let $\mathcal{S}^{(n)}$ be the feasible set of the convex program
solved at iteration $n$. Due to the updating rule in Algorithm \ref{algo:IterativeSDP}
(i.e., $\psi_{\mathtt{D}_{i}}^{(n+1)}=t_{\mathtt{D}_{i}}^{(n)}/\beta_{\mathtt{D}_{i}}^{(n)}$),
follows that $F(t_{\mathtt{D}_{i}}^{(n)},\beta_{\mathtt{D}_{i}}^{(n)},\psi_{\mathtt{D}_{i}}^{(n+1)})=f(t_{\mathtt{D}_{i}}^{(n)},\beta_{\mathtt{D}_{i}}^{(n)})$.
Similarly, we have $G\bigl(x_{\mathtt{U}_{j}},\mathbf{Q},\mathbf{q},x_{\mathtt{U}_{j}}^{(n)},\mathbf{Q}^{(n)},\mathbf{q}^{(n)}\bigr)=-g(x_{\mathtt{U}_{j}}^{(n)},\mathbf{Q}^{(n)},\mathbf{q}^{(n)})$.
This means that $(x_{\mathtt{U}_{j}}^{(n)},\mathbf{Q}^{(n)},\mathbf{q}^{(n)})\in\mathcal{S}^{(n+1)}$
and thus $u^{(n+1)}\geq u^{(n)}$ where $u^{(n)}$ is the objective
of \eqref{eq:Problem:Epi} at iteration $n$. The convergence proof
to a solution that satisfies KKT conditions follows the same steps
from \eqref{eq:proof:algorithm1} to \eqref{eq:stationary2:origin}
presented above.

\end{document}